\newif\ifusenix
\newif\ifacm
\newif\ifmcom

\usenixfalse
\acmtrue
\mcomfalse

\ifusenix
	\documentclass[letterpaper,twocolumn,10pt]{article}
	\usepackage{usenix2019_v3}

	\usepackage{times}
\fi
\ifacm
  \documentclass[sigconf,10pt]{acmart}

\usepackage[english]{babel}
\renewcommand\footnotetextcopyrightpermission[1]{} % removes footnote with conference info
\setcopyright{none}
\fi

\ifmcom
  \documentclass{sig-alternate-10pt}
\fi

\usepackage{xcolor}
\usepackage{placeins}
\usepackage{amsfonts}
\usepackage{balance}
\PassOptionsToPackage{hyphens}{url}\usepackage{hyperref}
\usepackage{color}
\usepackage{graphics}
\usepackage{graphicx}
\usepackage{url}
\usepackage{listings}
\usepackage{multicol}
\usepackage{multirow}
\usepackage[scaled]{helvet}
\usepackage{rotating}
\usepackage{xspace}
\usepackage{algorithm}
\usepackage{comment}
\usepackage{amsmath}
\usepackage{mathrsfs}
\usepackage{cancel}
\usepackage{textgreek}
\usepackage{enumitem}
\setlist{nosep,noitemsep,topsep=0pt,parsep=0pt,partopsep=0pt,leftmargin=*,parsep=0pt}

\usepackage{microtype}

\usepackage[capitalise]{cleveref}
\crefname{section}{Sec.}{Sec.} % type, single, plural. Tell cref to use abbr for sections.
\crefname{algocf}{Alg.}{Algs.}
\usepackage{titlesec}

\titlespacing\section{0pt}{6pt plus 4pt minus 2pt}{0pt plus 2pt minus 2pt}
\titlespacing\subsection{0pt}{6pt plus 4pt minus 2pt}{0pt plus 2pt minus 2pt}
\newcommand{\name}{BeamScatter\xspace}

\usepackage{graphicx}
\usepackage[font=small,labelfont=bf]{caption}
\usepackage[font=small,labelfont=bf]{subcaption}
\usepackage{multirow}
\usepackage{epsfig}

\newcommand{\DIFdel}[1]{}

\def\unnumfootnote{\xdef\@thefnmark{}\@footnotetext}

\newcommand{\bea}{\vspace{-2.0mm} \begin{eqnarray}}
\newcommand{\eea}{\end{eqnarray} }
\newcommand{\beas}{\vspace{-2.0mm} \begin{eqnarray*}}
\newcommand{\eeas}{\end{eqnarray*} }

\begin{document}

% Do not let latexdiff add border to figures
% \renewcommand{\DIFaddbegin}{}
% \renewcommand{\DIFaddend}{}

% Adjustments to equations must be set after \begin{document}
\setlength{\belowdisplayskip}{2pt}
\setlength{\belowdisplayshortskip}{1pt}
\setlength{\abovedisplayskip}{2pt}
\setlength{\abovedisplayshortskip}{1pt}
\setlength{\jot}{1pt}

% Adjustments to tables
\renewcommand\tabcolsep{2pt}
\renewcommand\arraystretch{1.2} % leave some space for math

\title{\name: Scalable, Deployable Long-Range Backscatter Communication with Beam-Steering} %Hierarchical Wake-up receivers for backscatter systems  }%for COTS Wi-Fi}

%\titlenote{Produces the permission block, and copyright information}
%\subtitle{Extended Abstract}

%\renewcommand{\shortauthors}{M.Dunna et.al}

%\author{Paper \# 325, 12 pages + 2 pages references}
 \author{{Manideep Dunna, Shih-Kai Kuo, Akshit Agarwal, Patrick Mercier, Dinesh Bharadia}}
% \{mdunna,s1kuo,akagarwa,pmercier,dineshb\}@ucsd.edu\\
%University of California, San Diego}
\affiliation{\institution{University of California, San Diego}}

% \authornote{Note}
% \affiliation{%
%   \institution{Affiliation}
%   \streetaddress{Address}
%   \city{City}
%   \state{State}
%   \postcode{Zipcode}
% }
% \email{email@domain.com}

%\keywords{Smart walls, Smart Surfaces, Virtual MIMO, Smart surfaces, WiFi Backscatter, Antenna design}
%\renewcommand{\shortauthors}{X.et al.}
\begin{abstract}
\noindent
WiFi backscatter tags can enable direct connectivity of IoT devices with commodity WiFi hardware at low power. However, most state of the art backscatter tag implementations in this area have limited transmitter to tag range, are not suitable to be deployed in a WiFi mesh network, and do not take full advantage of today's WiFi network capabilities such as MIMO. 
In this paper, we present \name, which can realize a backscatter tag based on MIMO techniques that can work at a very long separation of 28m from an access-point and enables their deployment in WiFi mesh networks. \name presents a novel technique to perform beam-steering on the MIMO tag at a very low power consumption of $88\mu$W while achieving a peak throughput of 500kbps. Next, \name creates a novel modeling framework to decide the optimal phase setting on the tag to steer the backscattered signal WiFi access point.
%Next, \name builds a novel hierarchical wake-up protocol, which, together with a custom ASIC, achieves a range of 30+ meters and the peak throughput of 1Mbps, with an average power consumption of $88\mu$W.

%and achieves sensitivity to wake-up at maximum optimal range of 30 meter from transmitter-to-IoT device.

% WiFi Backscattering can enable connectivity using existing infrastructure, 
% wherein an existing WiFi radio generates the WiFi signal, which is backscattered
% by an IoT device by encoding its data in the reflected signal, and this reflected signal
% is received by existing WiFi receiver, enabling IoTs to connect with WiFi at low power.
% However, most of existing work on Back-scattering WiFi/ambient signals has overlooked synchronization
% and accpeted limited range from transmitter-to-IoT device, as fundamental challenges. Even researchers have
% proposed to perform multi-hop backscattering to overcome limited transmitter-to-IoT device range~\cite{Varshney2020}. To 
% relax synchronization requirement, repetition in the IoT data bits is proposed, all leading stop-gap solution.
\end{abstract}

\maketitle

\section{Introduction}

%\todo{WiFi backscatter at a building level would be great, but needs to solve deployabliity}
Providing connectivity using existing wireless infrastructure in different environments, from smart homes and offices to smart stores and warehouses, simplifies the deployment of IoT devices like security cameras, human activity detection sensors in a factory building, etc. More recently, providing such connectivity via backscatter has become popular, as it allows for connectivity with  WiFi~\cite{ISSCC-backscatter,hitchhike,freerider,dunna2021syncscatter,wifibackscatter} while consuming micro-watts of power. 
These backscatter tags can be deployed at a large scale in big buildings to monitor conditions like temperature, air quality using temperature sensors, humidity sensors to enable building scale-sensing. 
To do so, we need to able to deploy an IoT sensor where ever needed in a building and should have full flexibility to choose a location for IoT sensor deployment. 
The key to achieve this objective is to make sure that backscatter links that connect the IoT devices to the WiFi network are not a hindrance while making the deployment decisions.

%\todo{don't where this paragraph goes}
%There have been on-going efforts in bringing WiFi to IoT space by few companies~\cite{haila,jeeva} to provide IoT connectivity using WiFi backscatter tags. These backscatter tags use ambient WiFi signals impinging on them instead of generating their own RF signals to achieve low power consumption. Low IoT device power consumption enables long sensor battery life\cite{IoT_power} and reduces the battery replacement costs\cite{IoT_battery} of IoT devices.

\begin{figure}[t!]
    \centering
    \includegraphics[width=0.9\linewidth]{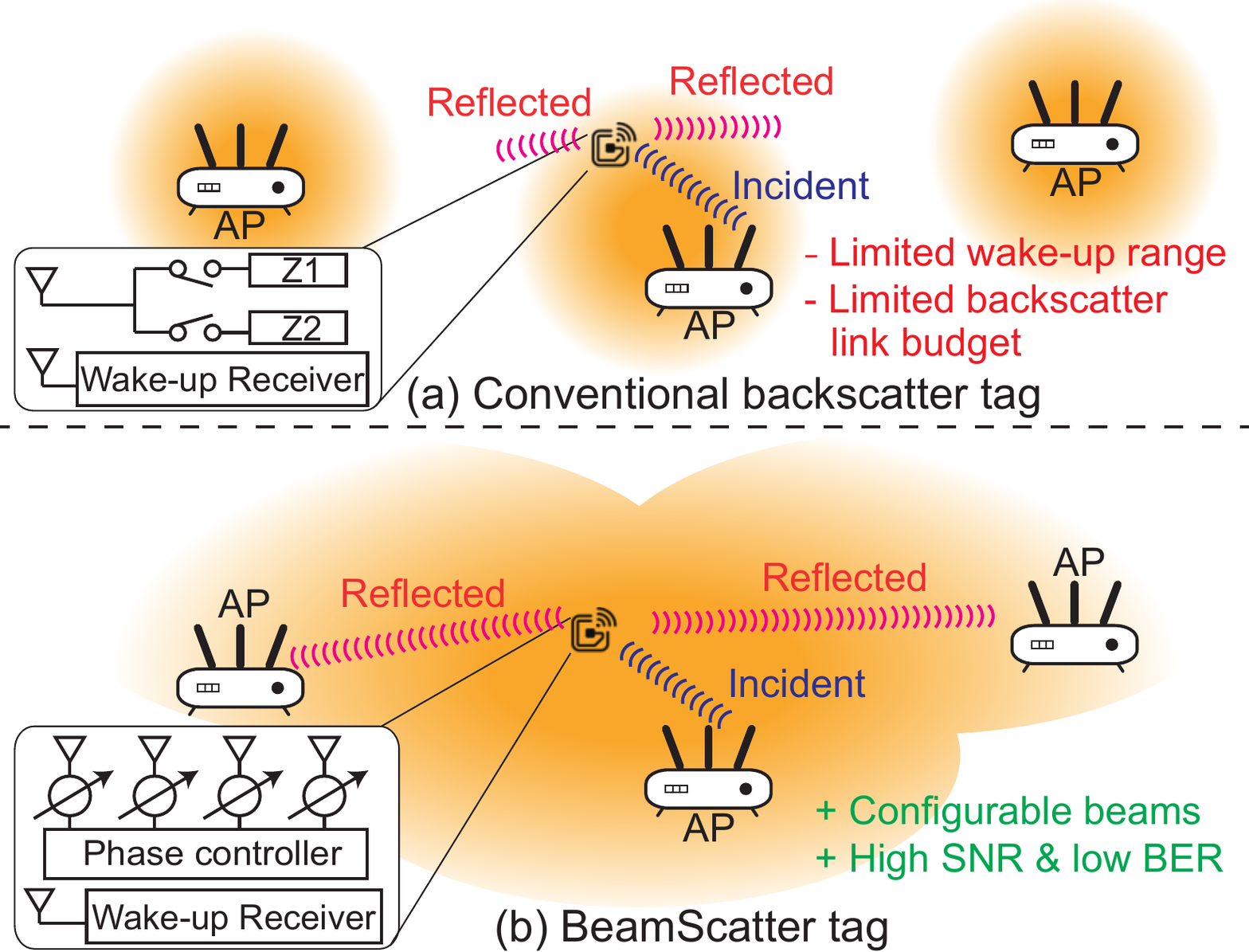}
    \caption{{Figure shows deployment challenges due to limited range for conventional backscatter tags with single antenna and is impractical to work in a network of WiFi APs. It shows the \name tag can steer the backscatter signals towards a receiving access point making the deployment of \name tag feasible at any location in a WiFi mesh network.}}
    \label{fig:intro}
\end{figure}

%\todo{Shihkai is helping with this now}

%\todo{ Before talking about deploy-ability problems... we explain first the operation? }
%However, previous backscatter works\cite{hitchhike,freerider,wifibackscatter,MOXcatter,liu2020vmscatter} cannot meet the deployability requirements in a large building. Before we explain about deployability, let us understand the operation of a WiFi backscatter tag. A WiFi backscatter communication link works as shown in Fig.~\ref{fig:intro}a. A WiFi access point sends a WiFi packet that acts as an excitation signal to the tag. If the tag detects the excitation signal, it wakes up and backscatters a WiFi packet on a different WiFi channel by embedding the IoT sensor data into the re-radiated transmission. Next, another another WiFi AP, part of the WiFi network listens to the backscattered WiFi packet and decodes the data, thereby enabling wireless internet connectivity to the IoT device. 

However, current backscatter tags\cite{hitchhike,freerider,wifibackscatter,MOXcatter,liu2020vmscatter} cannot meet the deployability requirements in a large building primarily due to range constraints. Due to range constraints, the end-user have limited freedom and they are forced to deploy the tags close to the access points. To understand why they have low range, let us understand how a WiFi backscatter tag works. A WiFi backscatter communication link works as shown in Fig.~\ref{fig:intro}a. A WiFi access point sends a WiFi packet that acts as an excitation signal to the tag. If the tag detects the excitation signal, it wakes up and backscatters a WiFi packet on a different WiFi channel by embedding the IoT sensor data into the re-radiated transmission. Next, another another WiFi AP, part of the WiFi network listens to the backscattered WiFi packet and decodes the data, thereby enabling wireless internet connectivity to the IoT device.

Backscatter tag deployability is affected firstly because of the high two-way pathloss experienced by a wireless signal: once in the downlink while going from the transmitter to the tag and again in the uplink while going from the tag to the AP. The high pathloss decreases the power of the backscattered signal reaching the AP and makes decoding of tag data difficult. Secondly, since a backscatter communication needs two APs or devices, the distance between transmitting and receiving AP influences the pathloss. Moreover, we do not have control over the AP to AP separation and the access point placement and separation is very specific to a floor plan decided based on the surveys\cite{WiFi_surveys}. The key to solving deployability problem is to design backscatter tags with large range. A large tag range allows a tag to be placed farther from an access point, increasing deployment flexibility. Also their design must easily scale for different access point separations to account for the variance in the access point densities across WiFi networks.

Past works\cite{hitchhike,interscatter} deal with pathloss problem by keeping the backscatter tags closer to an access point. However, they don't completely solve the deployability because of the limitation in choosing a tag's location. A simple way to decrease pathloss without compromising range is by amplifying the re-radiated signal from the tag. A class of backscatter works\cite{varshney2019tunnelscatter,amato2018tunneling} design a long range tag by creating high gain reflection amplifiers at low power consumption using a tunnel diode. However, the tunnel-diode approach fails with WiFi signals (20 MHz bandwidth) since tunnel diodes have limited bandwidth(a few 100kHz) over which they can provide amplification with good gain.

Our key insight in designing new backscatter tag architecture is that if we somehow introduce directional gain towards the excitation signals, we can reduce the backscatter link path loss and increase the tag range. To achieve this, we propose to use multiple antennas(MIMO tag) on the \name tag and do directional communication from backscatter tag to the access point. The intuition here is that multiple antennas on the tag capture more incident signal power from an access point and increase the backscattered signal power by beamforming the reflected signal towards WiFi access point, as shown in figure\ref{fig:intro}b. By controlling the number of antennas on the tag, we can deploy the tag anywhere in a network and scale it across different WiFi networks as well.

%If we have $N$ antennas on the tag, we can capture $N$ times the incident signal power because of $N$ antennas and beamform the backscattered signal again to provide an additional $N$ times gain. In total, we increase the backscattered signal power by $N^2$ times(a factor $N$ due to the receive beamforming gain and another factor $N$ from the transmit beamforming gain) and increase the tag's range. 
%The short range of the previous works\cite{hitchhike,dunna2021syncscatter} is due to the large path loss in the backscatter link.

The first challenge in designing \name tags is to determine how to steer the re-radiated signals from a backscatter tag in order to realize the beamforming gain. We introduce a phase-shifting element at each antenna so that the backscattered signals from each antenna constructively interfere at the receiving access point. The key question here is how to achieve the fine-grained phase-shifting operation with low loss? RF phase-shifters\cite{HMC247} are known to be very lossy and if we use them on the tag, the directional gains achieved from multiple antennas will be lost in the phase-shifter losses. To avoid this problem, we generate a phase shift at an intermediate frequency (IF) at each antenna by using the channel shifting mechanism already done on the backscatter tags. To impart a relative phase at each antenna, we delay the shifting clock appropriately at each antenna. So, each antenna on the \name tag needs an RF switch driven by a programmable delay clock. 

The next challenge is to make \name tag comply with real-world WiFi access point layouts. A typical WiFi network in a building comes with multiple access points operating on different WiFi channels. When a transmitter on WiFi channel 1 sends an excitation signal, the re-radiated signal from the tag reaches multiple access points operating on WiFi channel 6 at different distances from the tag and in different directions. First the tag has to determine, the best AP among the multiple APs operating on channel 6. Then the tag has to adjust the phases at each of its antenna for steering the re-radiated signal towards that AP. These phases depend on the tag's orientation and location w.r.t. the transmitting and receiving AP direction and it is very cumbersome to adjust the tag phases manually for every location. For the tag's usage to remain practical, the tag must automatically find out the best phase settings on it to steer the backscattered signal towards the best receiver AP when it enters a WiFi network. To solve this problem, we propose a geometric model for the signal re-radiated from the tag and use 4 CSI measurements from the tag's reflections to simultaneously determine the best WiFi AP to re-radiate the signal and the corresponding phase settings on the tag to do so.

\name is implemented and prototyped on a low-power CMOS integrated circuit\cite{ISSCC22-Shihkai} consuming only 88 $\mu$W. Furthermore, we have developed a discrete components-based \name tag to open source and provide the design to the community, which closely mimics the RFIC functionality but consumes significantly higher power compared to the RFIC version. We evaluate the tag in a large outdoor arena and show that the \name tag results in 56 meters with the RFIC version (35 meters from Tx-to-tag, 23 meters tag-to-Rx). The range is reduced for the discrete versions due to the higher insertion loss of the discrete RF switch. However, both versions demonstrate a 12 dB improvement over a single antenna tag. For an AP to AP separation of 40m, \name provides coverage of 14000 sq. ft, a 4x improvement over the state-of-the-art single-antenna backscatter designs. Consequently, we show that \name tag deployment requires only 4 APs to cover an entire warehouse of 14000 sq. ft as compared to hitchhike tag that need a dense deployment of 20 APs in the same area.

\section{Background and Motivation}
\begin{figure*}[t!]
    \centering
    \includegraphics[width=\linewidth]{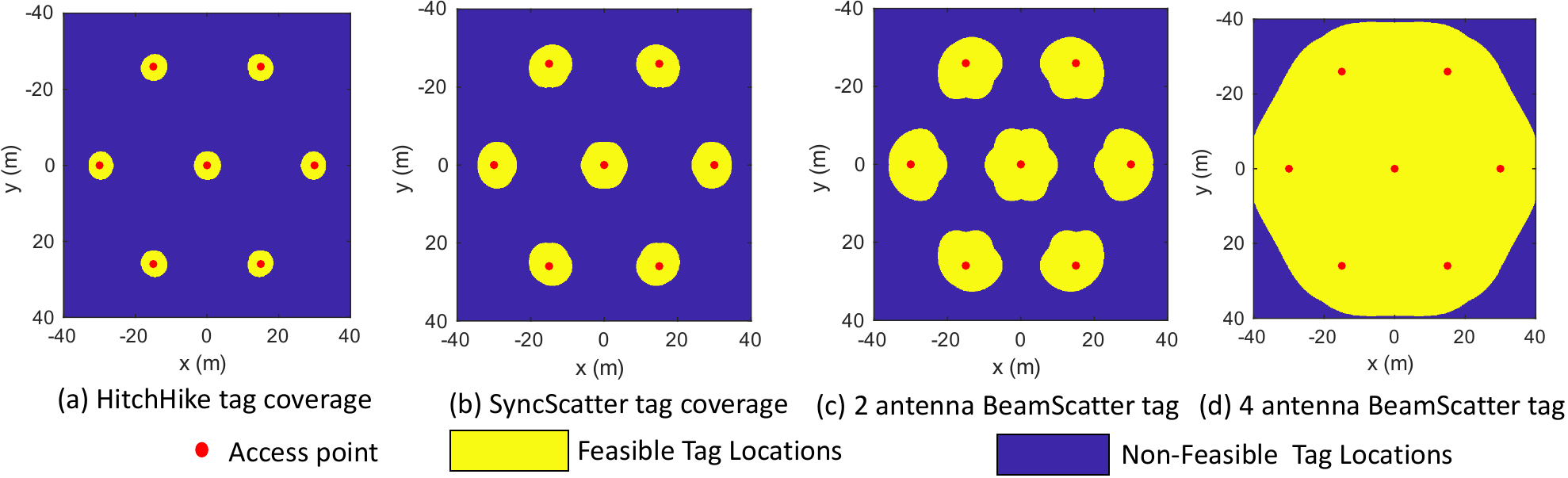}
    \caption{BeamScatter provides higher coverage than traditional hitchhike~\cite{hitchhike} or syncscatter~\cite{dunna2021syncscatter} tags.}
    \label{fig:coverage}
\end{figure*}

In this section, we will start with overview of bacskcattered signal power in a WiFi backscatter link. Then we proceed to explain the challenges with deployment of existing backscatter tags\cite{hitchhike,dunna2021syncscatter} and finally explain how a MIMO tag resolves this problem. 

%\textbf{Compliance with Access point cell planning:} As explained previously, to communicate with a backscatter tag, we need two WiFi access points operating on two  non overlapping WiFi channels in a WiFi network. Although at first glance, it might seem like a waste of resources to deploy two access points operating on two different WiFi channels to enable backscatter communication, that is not the case. This is because two neighboring WiFi access points in a network are by default assigned to operate on non-overlapping WiFi channels to minimize co-channel interference created by the access points. 

%\noindent
%\textbf{Protocol compliance:} Backscatter tags by themselves cannot initiate communication with an access point to transmit data. Hence, it needs the help of an AP to initiate a communication link. So, an AP sends instructions via downlink to get the tag ready to transmit its data.  For the tags to be compatible with Wi-Fi access points, backscatter tags must be able to interpret the AP's instructions. To achieve this, the tag is designed with a downlink receiver to understand the AP's instructions. Similarly, since the tag cannot actively emit WiFi signals on its own, it relies on the AP's transmissions to embed IoT sensor data on Wi-Fi signals. For the tag to operate at low power, the tag hardware has to be designed carefully while enabling both uplink and downlink. 

%\noindent
\textbf{Feasible tag locations}: Here, we explain the backscattered power from a tag as a function of its location and how to determine if a tag can be deployed at a location. When a Wi-Fi transmitter sends an excitation signal to a tag at a distance d1 from the transmitter, the signal undergoes attenuation proportional to the square of distance d1. When this incident power is more than the downlink sensitivity of the tag, the tag identifies the presence of an excitation signal. Then the tag reflects this signal which reaches another Wi-Fi access point at a distance d2 from the tag. The re-radiated signal also follows the inverse square law while propagating and the final received signal at the Wi-Fi access point varies inversely proportional to the product of the square of $d_1$ and $d_2$. For a tag location to be feasible, product $d_1*d_2$ must be such that the received signal power is above the Wi-Fi receiver sensitivity and  $d_1$ must be such that the incident signal is above the downlink sensitivity. 

%\noindent
\textbf{Challenges in real deployment:} Past backscatter works ~\cite{hitchhike,interscatter} like Hitchhike support a 50m backscatter range but they have a limited downlink range, so the tag can respond to the excitation signals only when the tag is within a 6m radius from an access point. When two neighbouring access points are deployed more than 12 m away from each other, there will be many infeasible tag locations due to the limited downlink range. A subsequent work, Syncscatter\cite{dunna2021syncscatter} improves upon the downlink range to 30m from the transmitter. So, now the backscatter tags are sensitive enough to detect the excitation signal even 30m away from the tag. Still, now the reflected power from the tag(depends on the product $d_1*d_2$) limits the tag's location from the access point to 10m. So, if the access points are located farther than 20m from each other, the tag cannot be deployed flexibly at an arbitrary location. Figure \ref{fig:coverage}a,b compares the coverage plots of hitchhike and syncscatter tag implementations. As can be observed, the syncscatter work increases the portion of coverage area for the tags. But still, it is not enough to cover the entire area using these tags.

\subsection{Towards flexible placement of tags} As discussed previously, the two-way backscatter pathloss is the main reason for the limited range of the backscatter tags. For flexible tag placement, the backscatter path loss has to be reduced. A straightforward way to reduce the backscatter pathloss is to use a directional antenna with higher gain on the backscatter tag. The higher transmit and receive gain at the antenna increases the strength of the re-radiated signal, and the signal can now reach an access point located farther. However simple this solution might look, using a higher gain fixed beampattern antenna comes with the following disadvantages. First, a fixed beampattern antenna requires manual adjustment to perfectly align the antenna towards the access points and this alignment changes with the tag's location. It becomes cumbersome to adjust the antenna's orientation everytime depending on tag location. Second, high gain antennas come with narrow beamwidth and so even a minor misalignment do not provide full directional gain and lowers the tag range. 

\textbf{Rethinking multi-antenna backscatter with \name:}  We propose an alternative approach to using directional antennas and Our insight is that as long as we can generate a custom configurable beam pattern that mimics high gain towards the transmitter and receiver access points, the solution is flexible enough to address the variability of the tag's location. Inspired by the concept of configurable antenna arrays, we propose to use multiple antennas on the tag to selectively receive and redirect backscattered signals. If we have $N$ antennas on the tag, we can capture $N$ times the incident signal power because of $N$ antennas and beamform the backscattered signal again to provide an additional $N$ times gain. In total, we increase the backscattered signal power by $N^2$ times(a factor $N$ due to the receive beamforming gain and another factor $N$ from the transmit beamforming gain). Figure \ref{fig:coverage}c shows the increased coverage from using two antenna \name tag compared to other approaches. For this specific access point separation of 30m, two antennas are not enough on the beamscatter tag and by using 4 antennas on the tag, it is able to cover the whole area. In the next section, we describe the design of such a steerable multi-antenna backscatter tag while maintaining a large downlink range and protocol integration. By building on prior work and expanding the range and flexibility of tag placement, \name pushes the needle on practical backscatter use.
% !TEX root = main.tex

\section{\name Tag Design Overview}\label{sec:design} 
\noindent
In this section, we start by describing the wake-up receiver to enable downlink to the tag and how to achieve redirectivity on the backscatter tag and the changes that need to be made on \name tag compared to a traditional backscatter tag. Then we move on to explain the algorithm to dynamically configure the backscatter tag to steer the incident signal towards a receiver AP. Finally, we explain the MAC protocol that allows the access points to coordinate with the backscatter tags using the wakeup receiver\cite{dunna2021syncscatter} on the tag and configure the tag. 

\subsection{Wakeup Receiver}
A backscatter tag is supposed to consume low energy by turning on its RF front-end only when a signal needs to be reflected. To let the tag know when it has to wake up the RF-front end, a tag has a special circuit called a wakeup receiver that is always powered on and continuously looks if there is an excitation signal incident on the tag. A tag is pre-configured to look for a special ON-OFF pattern in the incident signal. When a WiFi transmitter sends a special ON-OFF pattern using Wi-Fi packets specific to the tag, it detects this pattern and powers up the RF-front end to backscatter a WiFi packet.  

A wake-up receiver is implemented using an envelope detector followed by a comparator that tracks the changes in the power of the RF signals incident on the tag. The comparator's output is fed to a state machine to detect the pre-configured wake-up pattern. The wakeup pattern has a time duration of 10s of us. Hence the envelope detector can be designed to operate at a very low bandwidth and consume <5 uW power which can always be powered on. The wake-up receiver achieves a sensitivity around -35 to -40 dBm and is robust to collisions from other simultaneous Wi-Fi transmissions. We explain how we use the wakeup receiver to send downlink messages to the tag and also describe our implementation of the wake-up receiver in future sections.

% \todo{
% \begin{itemize}
%     \item Wakeup receiver is always on with low power consumption
%     \item It helps to wakeup the tag and put the tag in scannning modes to cycle through configurations.
%     \item It has a window based packet detecting state machine to enable downlink from the AP to the tag.
%     \item 
% \end{itemize}}

\subsection{Redirectivity design}
%-- ISSCC 
%\todo{Highlight $N^2$ gain here by explaining both transmit and receive beamforming gain.}
\noindent
Here, we describe how to design a backscatter tag that can redirect the incident signals on the tag towards a direction of interest. The process of sending signals in a specific direction is known as beamforming and is well studied as to how to enable beamforming on a wireless transmitter. Here, we develop an approach to perform beamforming of signals at the backscatter tag. Before going into the details of beamforming at tag side, let us first look at how beamforming is done at a transmitter. 

Transmit beamformers contain an array~\cite{phased_array_book} of antennas that are fed by signals that are phase-shifted with respect to each other. The phase difference between signals going to each antenna in the array determines the beamforming direction. For instance, consider a linear array of N antennas on a wireless transmitter with $\delta$ phase shift between signals fed to consecutive antennas. The signal emanating from different antenna has to travel different path-lengths that depends on the direction of travel and consequently signals from different antenna acquire different phase. The signal due to the array at an arbitrary point in the space is equal to the sum of the signals radiated from each antenna. So,the total signal in the direction $\theta$ is proportional to $\sum_{k=0}^{N-1} e^{jk\delta} e^{-j\frac{2\pi}{\lambda}kd sin(\theta)}$ where $d$ is the inter-antenna spacing and $\lambda$ is the signal wavelength. In order to steer the signal in a direction $\theta_0$, the wireless signals radiated from each antenna has to constructively add up i.e. they should all acquire the same phase in the direction $\theta_0$. By choosing the phase shift $\delta$ equal to $\frac{2\pi}{\lambda}d sin(\theta_0)$, we can add the signals from all the antennas constructively and steer the signal towards the direction $\theta_0$.  

\begin{figure}[t!]
    \centering
    \includegraphics[width=0.9\linewidth]{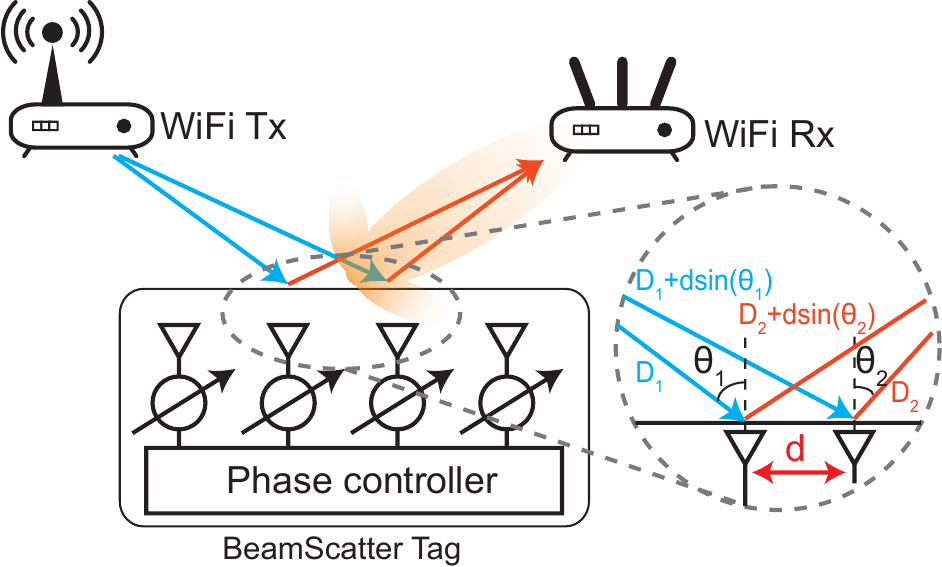}
    \caption{Shows the beam steering capability of the tag when the WiFi Tx and Rx AP are not in straight line with the tag.}
    \label{fig:redirectivity_illustration}
\end{figure}

Following the similar principle of constructive addition of signals by imparting an extra phase on each antenna, we build a multi-antenna backscatter tag to steer the backscattered signal towards a direction of interest. To understand this, consider a transmitter and a backscatter tag with $N$ antennas linearly spaced on it with inter-element spacing $d$ as shown in Fig~\ref{fig:redirectivity_illustration}. The signals arriving at each antenna travels a different path-length and acquire a different phase at each antenna. Also, each antenna on the backscatter tag can impart its own arbitrary phase on the signals before they re-radiate the incident signals. When each of these antennas reflects these incident signals, the signal strength at an arbitrary point in space depends on the phases imparted by the antenna and also on the angle at which the signal is incident on the tag. This is in contrast to the transmit beamforming where the signal strength depends only on the phases imparted by the antenna element.

To understand this, let the incident signal onto the backscatter tag makes an angle $\theta_1$ with its normal. Then the phase of the incident signal at each antenna is expressed in vector form as: $-\frac{2\pi}{\lambda}d sin(\theta_1) [0\  1\  2\  ... (N-1)]^T$. Let the additional phases applied at each antenna be $\Delta[0\ 1\ 2\ ... (N-1)]^T$. So, the total re-radiated signal at a point in an arbitrary direction $\theta$ is proportional to $\sum_{k=0}^{N-1} e^{jk\Delta} e^{-j\frac{2\pi}{\lambda}k [d sin(\theta) + d sin(\theta_1)]}$ which also includes the phase acquired by the re-radiated signal travelling in the direction $\theta$. Note that we assume that the tag is in the the far field\cite{far-field} of an access point while deriving the incident and re-radiated signal phases at each antenna. To steer the backscattered signal in the direction $\theta= \theta_2$, we have to choose $\Delta$ such that the re-radiated signals from all the antennas constructively interfere in the direction $\theta_2$. This results in choosing $\Delta$ to be equal to $\frac{2\pi}{\lambda}d [sin(\theta_2) + sin(\theta_1)]$. Our key insight is that, the phase acquired due to the additional path length for the incident signal has to be compensated as well in order to steer the reflected signal.  

So far we have seen how to enable beamsteering at the backscatter tag by applying additional phase shifts at each antenna. In addition to steering the beam, we should also be able to isolate the reflected signal from the other signals that are arriving at the receiver. To see why this is the case, consider a WiFi transmitter, a backscatter tag and a WiFi receiver. The WiFi receiver receives signals in two paths, the first one(direct path) directly from the transmitter and the second path(reflected path) via the backscatter tag where both the signal paths occupy the same frequency spectrum. When the backscatter tag wants to convey some data, the receiver will not be able to understand the tag data unless the receiver is able to separate the direct path signal from the reflected path signal. To solve this problem, previous backscatter works \cite{hitchhike,ISSCC-backscatter} propose to modulate the backscattered signal by shifting the signal to a different frequency spectrum so that the receiver can separate the direct and reflected signals. The idea here is to shift the incident WiFi signals on one channel to a completely different WiFi channel. For example, when a WiFi transmitter transmitting on Channel 1 centered at 2412MHz, the tag modulates the incident signal such that the resulting signal is shifted to Channel 6 (centered at 2437MHz) which has no overlap with Channel 1. Recall that we also need to apply different phase shifts at each antenna to steer the backscattered signal. So, how do we do achieve Channel shifting in addition to the applying phase shifts at each antenna?

\begin{figure}[t!]
    \centering
    \includegraphics[width=0.6\linewidth]{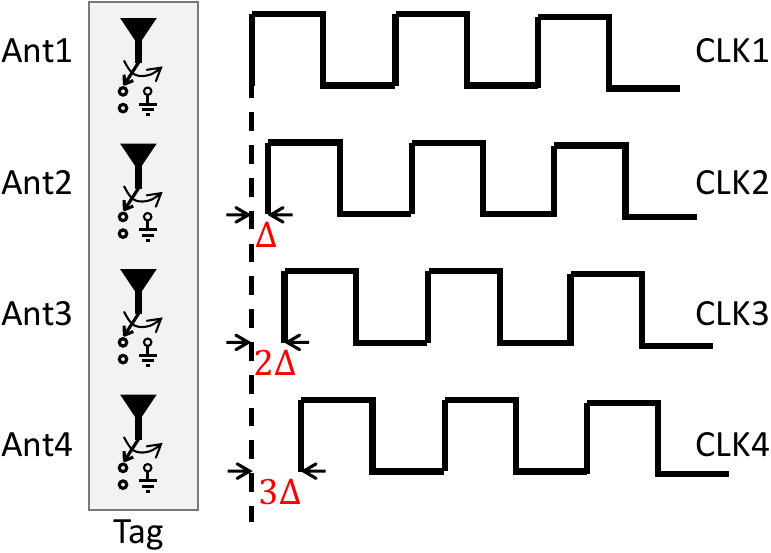}
    \caption{\name creates a variable phase shift in the clock at each antenna to achieve directional beamforming.}
    \label{fig:phase_per_antenna}
\end{figure}

To perform simultaneous channel shifting along with phase shifting at each antenna, we propose a novel structure on the tag by varying the phase of the switching clock driving the RF switches\cite{RF_switch} (Figure~\ref{fig:phase_per_antenna}) . Recall that the RF switch helps us in performing frequency translation. A clock drives each RF switch with a different phase that are tapped off a ring oscillator. The clocks' phase shifts are chosen so that the backscattered signal from the four antennas on the tag constructively adds up at the receiver. Half wavelength separates the antennas on the tag from each other to allow maximum beam steerability with given number of antennas.

\subsection{Determining tag's best phase setting}\label{sec:algorithm}
%\todo{algorithm}
\noindent
In this subsection, we discuss the algorithm to determine the best phase setting at each antenna element on the tag. Recall the notation that, $\Delta$ corresponds to the phase difference between two consecutive clocks modulating the RF switches. Here we consider the Tx, tag and AP setup as described in Fig~\ref{fig:redirectivity_illustration}. For a 4 antenna backscatter tag with $\frac{\lambda}{2}$ inter-antenna spacing, the signal strength at the receiving WiFi AP is proportional to  $\sum_{k=0}^{3} e^{jk\Delta} e^{-j\pi k [sin(\theta_2) + sin(\theta_1)]}$. Our goal here is to select the best possible $\Delta$ that decides the phase shifts at each antenna expressed in vector form as: $[0\  \Delta \  2\Delta \  3\Delta]^T$. 

Taking a closer look at the expression for signal strength at the receiver, we notice that when $\Delta$ is chosen to be $\pi[sin(\theta_2) + sin(\theta_1)]$, the signal strength is maximized. So, if we can somehow find the value of $\pi[sin(\theta_2) + sin(\theta_1)]$(let us denote it as $\alpha$ from now on), we can determine the optimal value of $\Delta$. For instance, if we know the transmitter and the receiver direction w.r.t. to the tag's normal, we can compute alpha but the problem with this approach is that we do not know the angles the Rx and the Tx makes when a new tag is introduced into the environment. Moreover, when the tag is attached to a mobile device, the tag needs to quickly adapt the phase shift values so that it can always steer the reflected signals towards the receiver. Here we resort to a different approach where we consider the channel state information (CSI) of the tag's received signal to find the best phase setting for the tag.

We devise a novel approach to compute alpha and in turn find the best $\Delta$ by performing just 4 channel measurements by setting the tag in 4 different phase configurations. In the first setting, we choose $\Delta = 0$ and record the CSI at the WiFi AP. Let us denote the CSI in this case to be h1. Similarly, we repeat the process by choosing $\Delta=\pi/2$, $\Delta=-\pi/2$, $\Delta=\pi$ and obtain the CSI at the WiFi AP for each of these phase settings. Let the signal strengths be h2,h3,h4 as given by equation \ref{eq:ratios}. Here $\beta_0,\beta_1,\beta_2,\beta_3$ refer to the random phase offset in each CSI measurement, $n_0,n_1,n_2,n_3$ is the noise in each CSI measurement and K is the pathloss factor due to the signal propagation in the air. 
\begin{align}
    h_0 &= Ke^{j\beta_0}(1 + e^{j\alpha} + e^{2j\alpha} + e^{3j\alpha}) + n_0\nonumber\\
    h_1 &= Ke^{j\beta_1}(1 + je^{j\alpha} - e^{2j\alpha} - je^{3j\alpha}) + n_1\nonumber\\
    h_2 &= Ke^{j\beta_2}(1 - e^{j\alpha} + e^{2j\alpha} - e^{3j\alpha}) + n_2\nonumber\\
    h_3 &= Ke^{j\beta_3}(1 - je^{j\alpha} - e^{2j\alpha} +je^{3j\alpha}) + n_3\label{eq:ratios}
\end{align}
Now we consider the ratio of the measured CSI $h_1,h_2,h_3$ with respect to $h_0$. To eliminate the effect of the random phases, we consider the magnitude of the ratios $\frac{h_1}{h_0},\frac{h_2}{h_0},\frac{h_3}{h_0}$. Now we compare these ratios to ideally what they would be in each of the cases. For example, the ratio of the signal amplitudes $\frac{h_1}{h_0}$ must ideally be equal to $r_1$ which is the ratio of the array factors for $\Delta=0$ and $\Delta=\pi/2$ as shown in \ref{eq:ideal_ratios}. Similarly, $\frac{|h_2|}{|h_0|}$ and $\frac{|h_3|}{|h_0|}$ must ideally be equal to $r_2$ and $r_3$ from equation \ref{eq:ideal_ratios}. 

\begin{align}
    r_1 &= \frac{(1 + je^{j\alpha} - e^{2j\alpha} - je^{3j\alpha})}{(1 + e^{j\alpha} + e^{2j\alpha} + e^{3j\alpha})} \nonumber\\
    r_2 &= \frac{(1 - e^{j\alpha} + e^{2j\alpha} - e^{3j\alpha})}{(1 + e^{j\alpha} + e^{2j\alpha} + e^{3j\alpha})} \nonumber\\
    r_3 &= \frac{(1 - je^{j\alpha} - e^{2j\alpha} +je^{3j\alpha})}{(1 + e^{j\alpha} + e^{2j\alpha} + e^{3j\alpha})} \label{eq:ideal_ratios}
\end{align}
Finally, we solve the optimization problem shown in equation \ref{eq:opt} to find $\alpha$ that minimizes the objective function. 
\begin{align}
    min \quad (|\frac{h_1}{h_0}| - r_1)^2 + (|\frac{h_2}{h_0}| - r_2)^2 + (|\frac{h_3}{h_0}| - r_3)^2 \label{eq:opt}
\end{align}
Intuitively, this objective function tries to keep the ratios of the measured CSI as close to the ideal ratios as possible. Unfortunately, there is no closed form solution to solve this problem. So, we solve this problem numerically by noting that the theoretical ratios $r_1,r_2,r_3$ are a function of $\alpha$ and they yield the same ratio for $\alpha = \alpha_0$ and $\alpha=\alpha_0+2\pi$. This implies the ratios are periodic with periodicity $2\pi$ and so it is enough for us to find a value of alpha in the range $-\pi$ to $\pi$ that minimizes the objective function. So, we evaluate the objective function over a grid of values in the range $-\pi$ to $\pi$ and find the value that minimizes the objective function. Once we determine $\alpha$, we set $\Delta = -\alpha$ on the tag to constructively combine and steer the signal towards the Rx AP.  Since in a WiFi network, there are multiple APs that could receive the backscattered signal, we estimate the best $\Delta$ for each of the Rx APs simultaneously using the 4 packet algorithm. Once the best $\Delta$ is found for each Rx AP, we set the tag with each of these best $\Delta$ phase and use the decoded data from the Rx AP that results in highest RSSI.
%Since we can calculate only the absolute value of $\alpha$, we have an ambiguity while deciding the correct value of $\alpha$. To eliminate the ambiguity, we can obtain the received signal powers for both  cases of $\Delta = \alpha$ and $\Delta = -\alpha$ and find which case results in good signal power to decide the optimal value of $\Delta$.  Here we have described the procedure to estimate the best $\Delta$ using received signal strength but in practice, we can get an estimate of the received signal strength from the Channel state information extracted at the receiver. 

%Let us say the tag has 10 possible steering configurations. Idea is to try each of the steering angle one after the other and see which steering angle gives the best signal power from the tag. 

% \begin{itemize}
%     \item 
%     \item Once the tag is in the scanning mode, the AP sends a trigger to the tag(downlink to the tag) to set the steering angle on the tag to the first configuration. Then the AP sends a small Wi-Fi packet to measure the backscatter signal strength for this  tag configuration. This process repeats until all the 10 configurations are tried.
%     \item Then the AP selects the best configuration and communicates to the tag.
% \end{itemize}

% \textbf{Downlink to the tag:}
% \begin{itemize}
%     \item AP needs to send three kinds of commands to the tag. First one is to put the tag in scanning mode. Second one is to change the tag's steering configuration to the next in sequence. Third one is to set the tag in backscattering mode to the best steering configuration.      
% \end{itemize}
\begin{figure}[t!]
    \centering
    \includegraphics[width=0.7\linewidth]{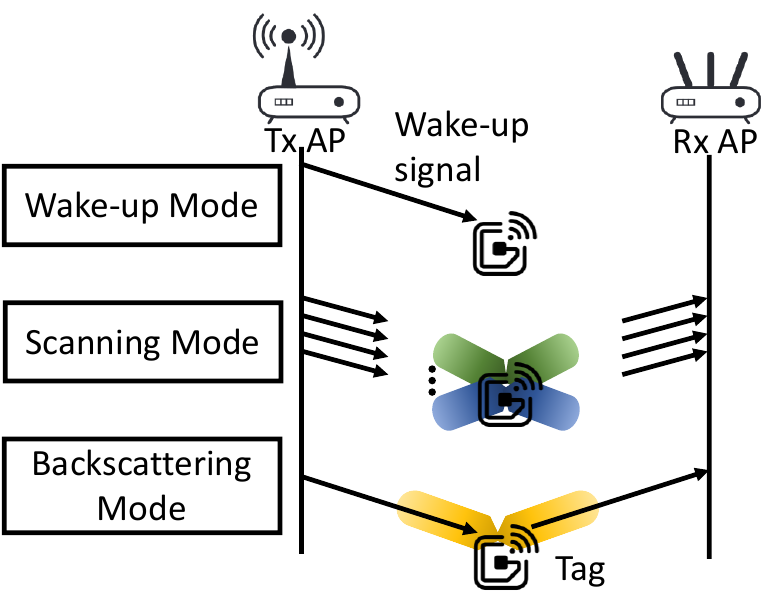}
    \caption{\name operates in three modes in coordination with the Tx AP: Wake-up mode, scanning mode, and backscattering mode.}
    \label{fig:mode_overview}
\end{figure}
\subsection{MAC protocol for AP-Tag Interaction:}
\noindent
In the previous sub-section, we have seen how to steer the backscattered signal from the tag by estimating the best possible phase settings on the tag. Here, we will see the protocol that makes it possible for the AP and the tag to interact with each other during this whole procedure. For the AP to interact with the tag, we utilize the Wake-up receiver\cite{ISSCC-backscatter,hitchhike} that is built onto the backscatter tag to send instructions from the AP to the backscatter tag which we refer to as the Downlink to the tag. Each instruction to the downlink wakeup receiver is accompanied by a CTS-to-self packet transmission to avoid collisions with the tag's instruction from surrounding WiFi devices. The tag is designed to operate in three different modes as shown in Figure~\ref{fig:mode_overview}. In the first mode, the tag listens for a wake-up pattern. Once the Tx AP sends the wakeup pattern, the tag detects it and is set in scanning mode. In the scanning mode, the tag loops through four phase settings ($\Delta=0,\Delta=90,\Delta=180,\Delta=270$). Once the tag enters the scanning mode, the AP sends instructions to the tag on the downlink to set the tag to first one in the list of four phase settings. Then the Tx AP sends a 802.11g WiFi packet. In beamscanning mode, the tag just shifts the frequency of the WiFi packet without encoding any tag data on it and Rx AP collects the CSI for that packet. Now the Tx AP sends a downlink instruction to the tag to change to the next phase setting and this process continues until CSI is measured for the list of four phase settings. Then the best phase settings for the tag are estimated at the Rx AP according to algorithm described in section\ref{sec:algorithm} and sent over to the Tx AP via the ethernet backbone. Next the Tx AP again sends information about the best phase setting to the tag via tag downlink and set the tag to beamsteering mode in which the tag data can be encoded into the incident WiFi packets. From now on, the tag is frozen in the best phase setting and the the Tx AP uses 802.11b packets as the excitation signals which are used by the backscatter tag for modulation. Here we assume that all the APs in the mesh network are connected to each other over Ethernet and they can coordinate with each other by exchanging information between them.
%!TEX root = main.tex
\section{Hardware Design and Implementation}
 \begin{figure}[t!]
    \centering
    \includegraphics[width=1.0\linewidth]{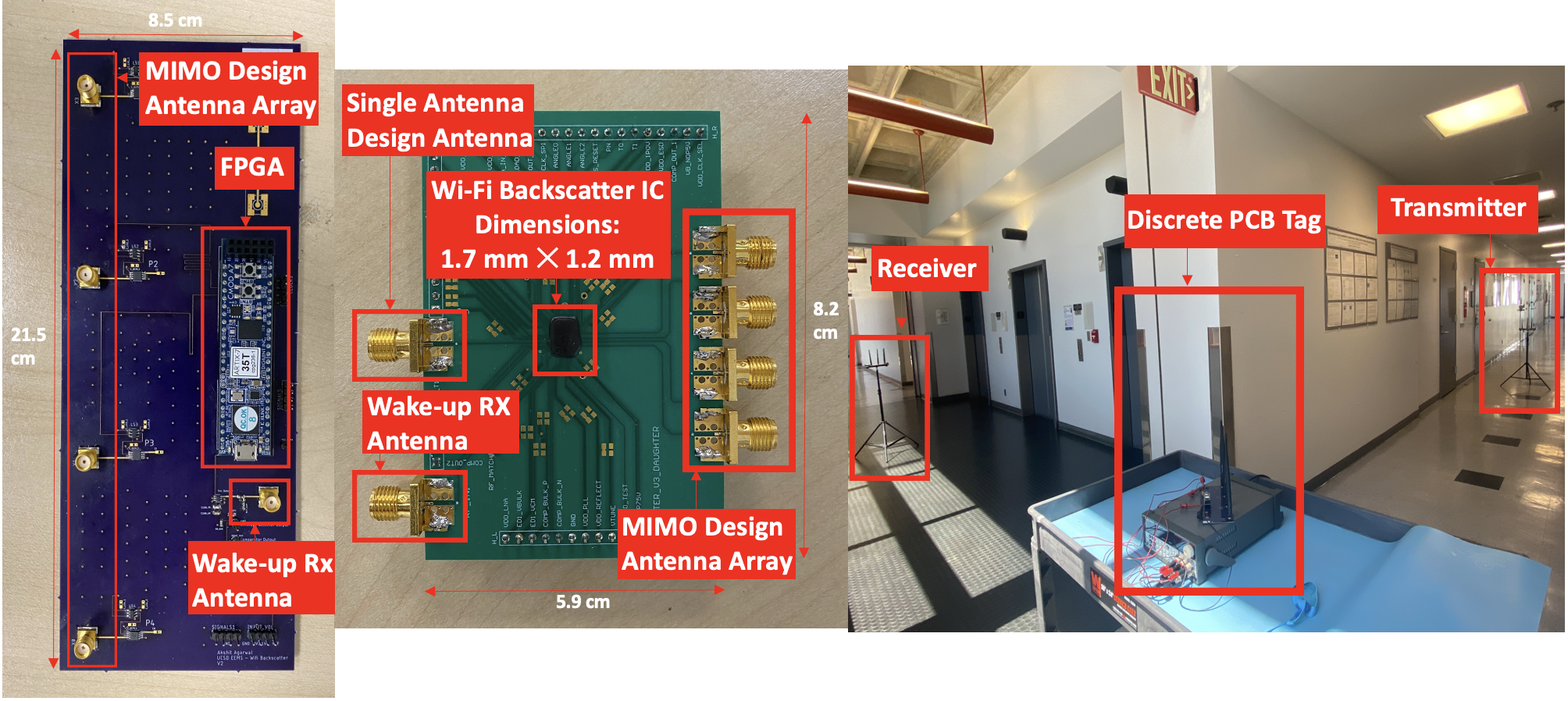}
    \caption{(a) shows the discrete components based \name tag design (b) RFIC version of \name's tag on a pcb (c) experimental setup showing the tag, WiFi transmitter and WiFi receiver}
    \label{fig:SETUP}
\end{figure}
\noindent
We have prototyped \name tag with a Custom RF Integrated circuit to develop ultra-low power consumption of the \name tag. RFIC version was fabricated in TSMC's 65nm GP process. Furthermore, to help the community, we have also designed a tag using off-the-shelf components, which mimics performance behaviour close to the RFIC in most aspects, but consumes a bit higher power. In the subsequent sections, we will go over the first the discrete components based tag design first, then discuss the RFIC version of our tag~\cite{ISSCC22-Shihkai}, with their individual blocks that make up the tag. Finally, we show the commercial AP used in our experiments and the firmware used with the AP. 
\begin{figure}[t]
\centering
\includegraphics[width=\linewidth]{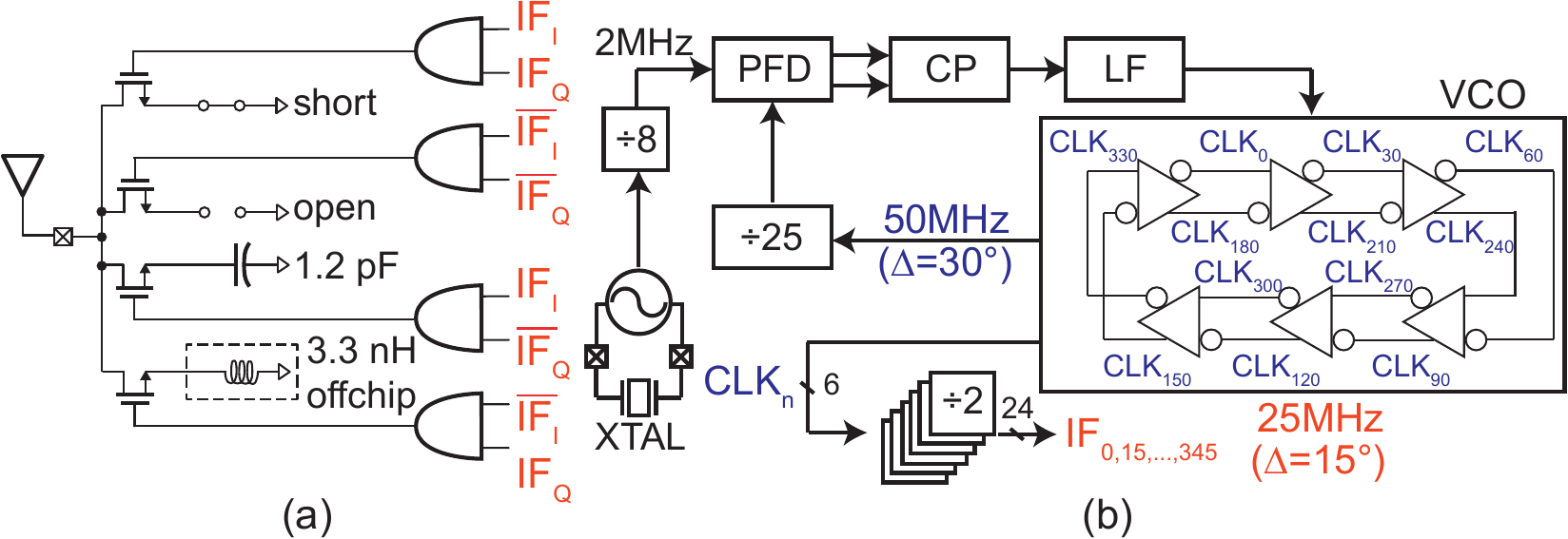}
%\subcaption{}

%\begin{minipage}{0.22\textwidth}
%\includegraphics[width=\linewidth]{figures/Reflector_3_crop.pdf}
%\subcaption{}\label{fig:reflect_no_splitter}
%\end{minipage}
%\begin{minipage}{0.22\textwidth}
%\includegraphics[width=\linewidth]{figures/PLL_divider_v2_crop.pdf}
%\subcaption{}\label{fig:logic_select_a}
%\end{minipage}
% \hfill
% \begin{minipage}{0.36\textwidth}
% \includegraphics[width=\linewidth]{figures/Selection_crop.pdf}
% \subcaption{}\label{fig:logic_select_b}
% \end{minipage}
% , \textbf{(c)} Clock selection logic. 
\caption{\textbf{(a)}   SP4T fully reflective radiators without power splitter, \textbf{(b)}PLL and divider blocks}\label{fig:PLL_reflector}
\end{figure}

% \begin{figure*}[t]
% \centering
% \begin{minipage}{0.24\textwidth}
% \includegraphics[width=\linewidth]{figures/Reflector_3_crop.pdf}
% \subcaption{}\label{fig:reflect_no_splitter}
% \end{minipage}
% \begin{minipage}{0.24\textwidth}
% \includegraphics[width=\linewidth]{figures/PLL_divider_crop.pdf}
% \subcaption{}\label{fig:logic_select_a}
% \end{minipage}
% % \hfill
% % \begin{minipage}{0.36\textwidth}
% % \includegraphics[width=\linewidth]{figures/Selection_crop.pdf}
% % \subcaption{}\label{fig:logic_select_b}
% % \end{minipage}
% % , \textbf{(c)} Clock selection logic. 
% \caption{\textbf{(a)}   SP4T fully reflective radiators without power splitter, \textbf{(b)}PLL and divider blocks}
% \end{figure*}
%\todo{PCB discrete setup}

\subsection{Discrete PCB based \name tag}
\noindent\textbf{\name modulator block} 
We implement a four-port antenna array on a printed circuit board (PCB) tag which is controlled by a CMOD A7 FPGA~\cite{cmod}, as shown in Fig. \ref{fig:SETUP}(a) The separation between each of the ports is fixed to 6.25 cm which is the half wavelength of 2.4GHz WiFi signals. The ports are connected to the RF outputs of four ADG902 RF switches~\cite{rfswitch}. The control signals for these switches are 50\% duty-cycled 25MHz phase-shifted clocks produced by the FPGA. The antennas connected to the RF output of the switches shift the incoming signal by 25MHz and change the phase of the signal.    

\noindent\textbf{Downlink, Wake up and Synchronization Receiver}
In our design, the \name tag is kept in low-power mode by turning it on only when there is a WiFi signal incident on the tag, with a specific sequence as well. So, \name tag must be capable of detecting the incident signal and for that purpose, we have implemented a wake-up receiver\cite{dunna2021syncscatter} that is always powered on and looks for if any excitation signal is incident on the tag. We have implemented the wake-up receiver using a wide-band envelope detector LT5534\cite{LT5534} that responds to 50MHz to 3GHz RF signals. The envelope detector outputs the envelope of the incident signal which is then passed to the comparator MAX49140\cite{MAX49140} to digitize the signal envelope. The digitized signal is then passed to an FPGA that samples the signal to find out if the signal envelope matches the wake-up pattern. The envelope detector can support synchronization to an accuracy below 150ns\cite{dunna2021syncscatter} with bandwidth >10MHz. Furthermore, our PCB design can implement hierarchical wake-up receiver similar to \cite{dunna2021syncscatter}. We leverage the comparator to develop complete downlink similar to~\cite{hitchhike}. We have designed special sequences followed by tag wake-up to convey the downlink data, or beam-scanning protocol and the optimal phase configuration.  

\noindent\textbf{Programmable delay generation per antenna with Clock Generation}
The CMOD A7 FPGA is used to generate 25 MHz phase-shifted clocks that act as the control signals for the RF switches. An IP core is used to generate four 150MHz quadrature clocks, which are then fed to a clock division module that outputs six 25MHz phase-shifted clocks for each of the input clocks. This gives us a total of twenty-four 25MHz clocks that differ in phase by 15 degrees each. These clocks are multiplexed such that the four switches receive the control clocks with phases 0, $\Delta$, 2$\Delta$ and 3$\Delta$ for $\Delta$  values in the set \{0 to 360 degrees, with 15 degree step size\}. %\todo{Double check: 30 degree or 15 degree in the set?}     
%\todo{Write about 0 and 180 phase changing}

\subsection{RF Integrated Circuit (RFIC) Design}
\noindent
Discrete components based version closely mimics the RFIC version of \name's tag, but the RFIC one shown in Fig. \ref{fig:SETUP}(b) is fully integrated and extremely power efficient.
The RFIC design includes a wake-up receiver, hierarchical synchronization/downlink receiver, single-side-band (SSB) single-antenna modulator, SSB MIMO beam-steering modulator as well, and an array of low-loss backscatter radiator switches. The fabricated chip is directly bonded to the testing PCB where external antennas are connected and power traces are routed. An off-chip 16MHz crystal oscillator provides the base-band clock for the chip. In \cite{dunna2021syncscatter} the wake-up receiver, synchronization receiver, and SSB single-antenna modulator are introduced, so this paper describes the proposed radiator switches and the SSB MIMO beam-steering modulator in the remainder of this section.

\noindent\textbf{Radiator}
To reduce the insertion loss, the fully-reflective radiator in \cite{meng202112} removes the absorbing loads in \cite{dunna2021syncscatter} and provides a 6dB insertion loss improvement. However, a lossy power splitter/combiner is still utilized to create I/Q paths. To get rid of the power splitter, in this work we propose to use a single SP4T switch radiator controlled directly by the I/Q clocks, where the I/Q combination is now happening in IF instead of RF, as shown in Fig.~\ref{fig:PLL_reflector}(a). With correct clock timing from the clock selection logic, four fully-reflective loads (open, short, capacitor and inductor) connected to the SP4T switch are cycled through based on the tag data to generate the SSB QPSK backscatter. With such architecture, 3dB lower insertion loss is achieved compared to \cite{meng202112}. To enable beam-steering, the proposed architecture in this paper leverages four copies of the radiator to form the MIMO array. 

\noindent\textbf{Clock Generation}
To generate multi-phase clocks, ring oscillator based voltage controlled oscillator (VCO) is used because clock with different phases can be easily tapped out from each stage. To ensure the phase stability, a phase-locked loop (PLL) shown in Fig.~\ref{fig:PLL_reflector}(b) is used to lock the output clock phase with the 2MHz reference clock coming from the 16MHz base-band crystal oscillator. A standard type-II analog integer-N PLL with divider ratio 25 is used to generate 50MHz clocks. The VCO in the PLL is made of a 6-stage differential ring oscillator, which produces clocks with 30-degree phase resolution. CLK \{0,30,60,90,120,150\} are further passed through divide-by-two blocks to generate 25MHz IF clocks with 15-degree phase resolution, which can frequency-shift the incident signal between adjacent advertising channels. The power consumption of PLL is 30$\mu$W.  

Before diving into the clock selection logic, we have to specify the requirement of final clocks fed to four antennas in the MIMO array: 1) To implement QPSK backscatter on a single antenna, four phases of quadrature clocks are required, which are chosen by a 2-bit tag data. 2) To further extend to SSB, instead of generating only one set of I-phase quadrature clocks, another 90-degree shifted Q-phase set of clocks is also required. 3) To support MIMO beam-steering, the above-mentioned I and Q sets of clocks should be phase-shifted to the individual antenna in the array based on the angle estimation. To fulfill the above mentioned requirements, a clock selection logic\cite{ISSCC22-Shihkai} is designed to multiplex a clock from the rign oscillator to each RF switch. The selection logic enables choice of clocks at each antenna with $15^{\circ}$ phase difference. In total, MIMO beamsteering selection logic consumes 59$\mu$W.

%Fig.~\ref{fig:logic_select_b}

%In the first layer of selection logic, 3-bit angle selection control signal from angle estimation logic determines the phase difference of clocks between antennas. For example, with $IF_{Qua}$ defined as a set of clocks \{$IF_0$, $IF_{90}$, $IF_{180}$, $IF_{270}$\}, when angle selection bits are 000 (minimum angle setting), the output of first selection layer will be $IF_{Qua}$, $IF_{Qua} +15^\circ$, $IF_{Qua} +30^\circ$ and $IF_{Qua} +45^\circ$, which have 15 degree phase shift between them.
%The output from the first selection layer is further fed to the second layer digital SSB IF mixers \cite{ISSCC-backscatter, dunna2021syncscatter}, which are 4-1 MUXs controlled by 2-bit tag data. By selecting one clock from the quadrature IF clocks, QPSK backscatter modulation can be achieved. Finally, the plus-minus control bit from estimation logic arranges the order of clocks that go to the MIMO array.  The MIMO beam-steering selection logic consumes 59$\mu$W in total.

%\todo{Bring up discussion on ring oscillator here saying that the clock generation with multiple phases comes up for free because of the structure of the ring oscillator.}
\subsection{\name tag vs Active tag}
As we have seen in the previous sections, the tag's power consumption increases due to multiple RF switches and multiple phased clocks generated on the tag. Here we will present why \name tag is a viable low power solution compared to an active tag. In \name tag, the tag needs to perform 4 rounds of downlink and uplink to find the best possible phase setting. Considering a worst case scenario, where the tag finds best possible phase every time it sends a payload,each round contains a wakeup pattern and a CTS-to-self packet together spanning 400$\mu$s which is followed by a WiFi packet payload. In the time duration corresponding to the 4 rounds, the tag spends $88\times400\times4 = 0.14\mu J$ of energy in addition to backscattering a payload. Comparing this to an BLE active tag transmitting at 1mW power level, as long as the payload is more than 1ms, the overhead energy consumption on \name tag is still 10 times lower than an active tag.

\subsection{Working with COTS WiFi AP}
\noindent
We implemented BeamScatter using the open-source Nexmon CSI Extraction tool\cite{nexmoncsi}, running on an ASUS RT-AC86U router. This tool utilizes modified firmware on the AC86U to read channel frequency response and RSSI estimates directly from the internal memory of the router's onboard Broadcom chipset. The Access point is set in monitor mode to sniff on all the packets over the air and we filter the packets used in our experimentation based on MAC address and extract the channel state information(CSI), shown in Fig.~\ref{fig:SETUP}(c). 

The host-pc mimicks the cloud in our case, is connected to both the transmitting and receiving APs,  which performs the decoding of the tag data. The transmitting WiFi AP sends the wakeup sequence for the tag, followed by beam scanning meassage. Upon which tag starts beam-scanning, the receiving WiFi APs run our algorithms which are implemented on the host-pc. The extracted CSI for the four packets is processed in near real-time to compute the optimal phase configuration. Following that the receiving APs learns the optimal phase configuration, the phase configuration are transmitted by transmitting WiFi AP to the tag via downlink, followed by backscatter uplink communication from the tag. The received data is collected via wireshark running on the APs, which then recovers the tag data.

%This tool is widely used to extract CSI and has shown to be viable for localization in the past\cite{ubilocate https://dl.acm.org/doi/10.1145/3458864.3468850}. 

%possibly also add this line if you want to give more info about this tool's uses, may be unnecessary:

%\input{implement}
% !TEX root = main.tex
\section{Evaluation}
 \begin{figure*}[t!]
    \centering
    \includegraphics[width=0.9\linewidth]{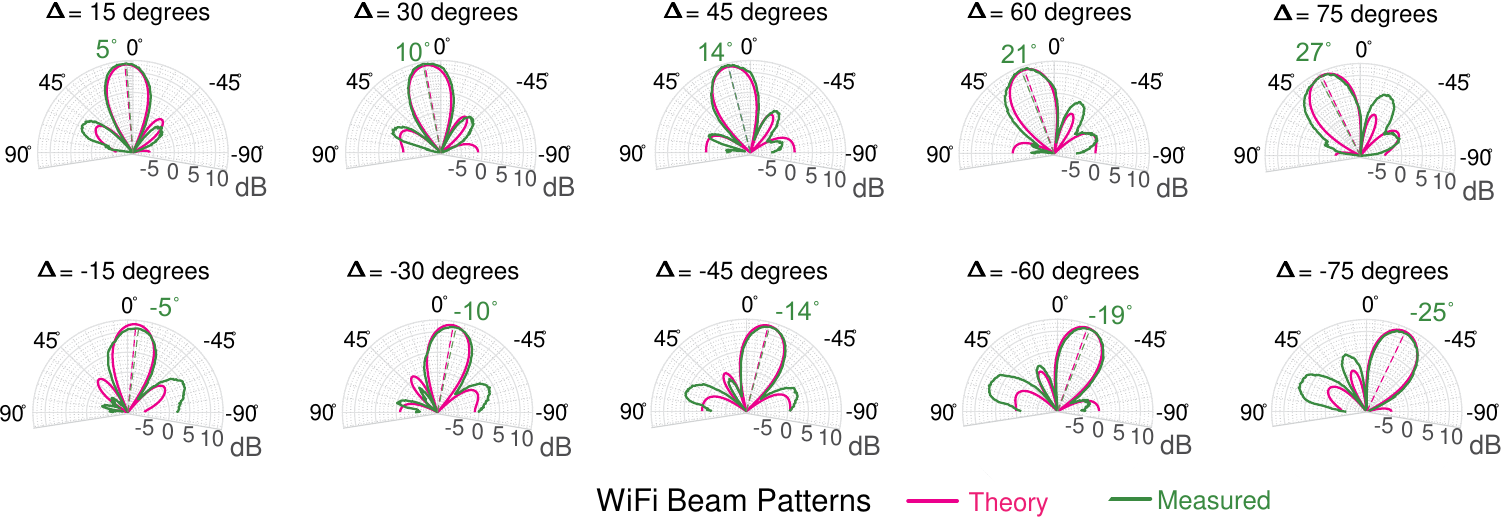}
    \caption{Beampatterns of the measured backscatter signal for different phase settings($\Delta$ denotes the phase difference between the consecutive antennas) when the excitation signal is incident normal to the tag . The measured beampatterns are compared to the theoretical beampatterns. The measured beam-steering angles match very closely to that of the theoretical value.}
    \label{fig:beampatterns}
\end{figure*}
\noindent
We evaluate \name tag in outdoor scenarios, to conform with Omicron-related policies. We conduct the experiments by placing the Transmitter AP, Receiver AP and the Tag at different locations in a large outdoor arena of 60m x 40m dimensions.
Furthermore, we construct both line-of-sight (LOS) and non-line-of-sight (NLOS) experiments for the \name tags placement. We first present microbenchmarks to demonstrate the working of the \name tag. We place the tag, Tx and Rx APs at different locations in an anechoic chamber to show that the tag is able to beamsteer the backscattered signals towards a receiving AP in every configuration. Then we evaluate the effectiveness of the proposed algorithm to estimate the best tag configuration and show its closeness to optimal. Finally, we provide end-to-end performance results in terms of goodput, BER, and range of the 4 antenna \name tag by placing the tag at 100 locations. Furthermore, we place both RFIC and discrete components based \name tag, and evaluate the performance for both. For each location, we conduct single antenna-based experiments as a baseline to contrast the improvements.

%\section{Microbenchmarks:}
%   \begin{figure}[!t]
%  \centering
%  \includegraphics[width=.9\linewidth]{figures/alpha_cdf_0deg.pdf}
%  \caption{CDF of estimated $\alpha$ when the Tx, Tag and Rx are in a straight line.}
%  \label{fig:alpha_cdf_0deg}
% \end{figure}

\subsection{Microbenchmarks:}
\noindent
\textbf{Beampattern Measurement:} Here we present a microbenchmark to demonstrate the beam-steering capability of the tag by measuring the beam pattern of the backscattered signals by doing a normal incidence of RF signals onto the tag in an anechoic chamber i.e. the incident signal at the input port of the antenna array has the same phase. For this measurement, the incident signal is normal incidence onto the tag by the following arrangement, which is then reflected in different directions. The horn antenna for TX and RX are placed to maximize the interference suppression to enable us to measure the beam pattern (1 meter from tag).  The reflected signal is shifted in frequency by 25MHz and its phase is modified at each port of the array due to the different phases of the baseband clocks that control the switches. When the reflected signals exit the antenna, they reach the receiver and combine to give a signal proportional to the magnitude of the beam pattern. Here we use a Spectrum analyzer to monitor the received signal power. Then we rotate the \name tag around its axis by mounting the PCB on a motor that has very fine angle control. We repeat this method for different tag configurations (different $\Delta$- phase shift between consecutive antennas) and plot the received signal power as a function of the tag rotation angle to obtain the measured beam patterns. We compare the measured beam patterns with the theoretical beam pattern as shown in Figure~\ref{fig:beampatterns} and those match closely. We observe a maximum gain of 11 dB close to the 12 dB theoretical gain(= 10$\times \log(16)$) for a four antenna array operating in backscattering mode. We note that the side lobes are slightly compared to the theoretical patterns arising due to the slight offset in the clock phases from the programmed values. 
 
%  \subsection{Rx Power variation for different tag phase configurations:}
%  -- To show that, the backscatter signal power is changing with different phase settings on the tag.
 
%  X-axis: phase difference between consecutive antenna 
%  Y-axis: Rx signal power value
%\textbf{RSSI comparison with a single antenna tag:}
\subsection{Measurement Case study:}
\begin{figure}[t]
\centering
\includegraphics[width=.9\linewidth]{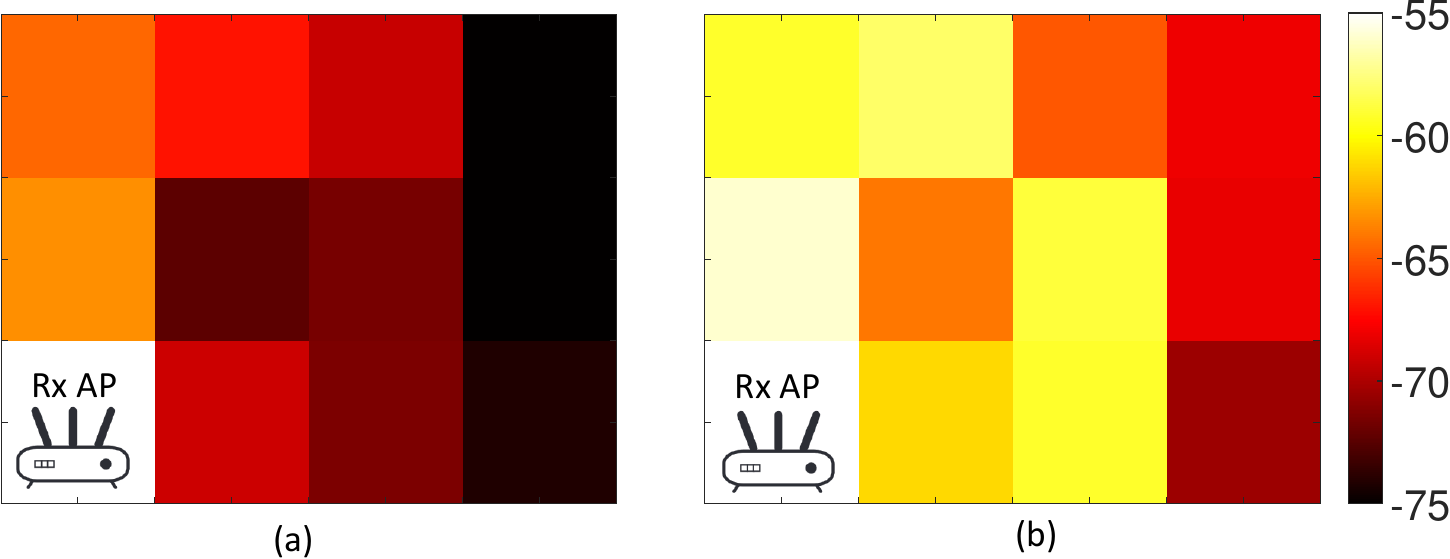}
\caption{Figure showing heatmaps of received signal strength at the Rx AP in a 4.5 m x 4 m room using a) single antenna(hitchhike) tag b)Beamscatter tag}
\label{fig:heatmap}
\end{figure}

We have done a measurement study in a 4.5 m $\times$ 4 m to show the RSSI improvement provided by \name tag. For this, we divided the room into a 4 x 3 uniform grid and placed the Tx and Rx AP at the corners of the room. Then at every grid point in the room, we placed a single antenna tag as well as four antenna \name tag and collected the RSSI measurements at the Rx AP for both the cases. Figure\ref{fig:heatmap} shows heatmap of RSSI at the AP for both SISO and MIMO tags. Across the tag locations, the RSSI has increased by a minimum of 4 dB and a maximum of 12dB. On an average, we observed 8 dB improvement across all the points.  

\subsection{Estimating the best tag configuration:}
\noindent
Next, we present the results for estimating the best phase setting on the tag. This experiment is conducted in an anechoic chamber where the transmitter sends Wi-Fi packets on WiFi channel 1 and multiple Wi-Fi receivers monitor the backscattered packets on WiFi channel 6 implemented using ASUS RT-AC86U access points loaded with Nexmon firmware\cite{nexmoncsi}. To show the effectiveness of the algorithm in estimating the best configuration, we conduct experiments by placing the Rx AP in different directions w.r.t. to the tag's normal while keeping the transmitter and three receivers approximately 1m away from the tag in the anechoic chamber. For this experiment, we collected the CSI and RSSI measurements by transmitting 802.11g WiFi packets for different phase shift values $\Delta$ on the tag and show the RSSI improvement provided by the multi-antenna backscatter tag. We used 4 antennas on the tag and arranged the tag to have normal incidence of the signal from the transmitter.
Then we use the CSI measurement at each receiver for 4 different tag phase configurations as described in section \ref{sec:algorithm}, and solve for the best phase setting to steer the reflected signal for each receiver location. We repeat the estimation procedure for optimal $\Delta$ using multiple CSI measurements to observe the variation in the estimates.  

%and plot the CDF of the estimates in Figure~\ref{fig:alpha_cdf_0deg}. The sharp CDF plot indicates that the algorithm is consistently(around 85\% of the time) estimating the correct value of $\alpha$. We also compare the RSSI upon setting the tag in the estimated best configuration and compare it with the true best setting.

\begin{figure}[t]
\centering
\begin{minipage}{0.23\textwidth}
\includegraphics[width=0.8\linewidth]{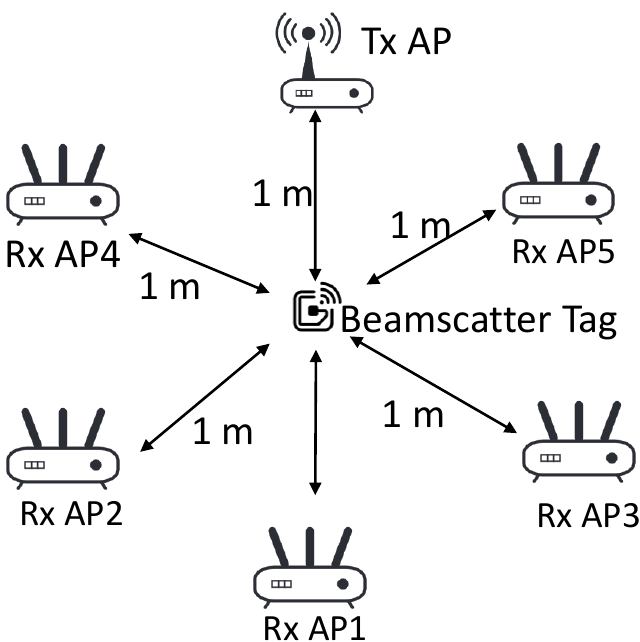}
\subcaption{Anechoic chamber setup of Tx, tag and multiple Rx APs}
\label{fig:multi_ap_setup}
\end{minipage}
\begin{minipage}{0.24\textwidth}
\includegraphics[width=\linewidth]{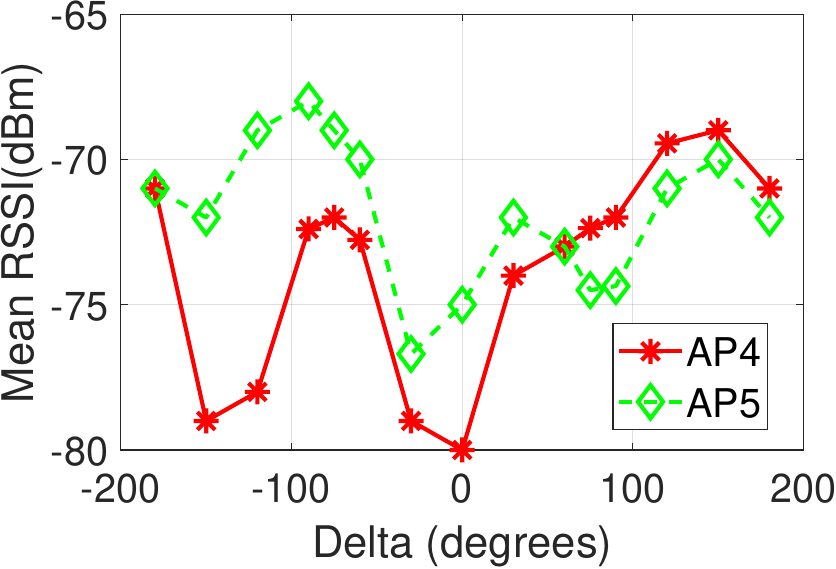}
\subcaption{Mean RSSI for Rx APs placed between Tx and tag.}
\label{fig:rssi_location}
\end{minipage}
\caption{Figure showing the placement of Tx, tag and Rx AP in multiple configurations and the recieved signal strengths}
\end{figure}

\noindent
\textbf{Effect of Rx Access Point's location:}
In real deployments, we do not have control over the placement of the transmitter and receiver relative to the tag's location, and they can be on the same side of the tag or on the opposite sides of the tag as shown in Figure\ref{fig:multi_ap_setup}, depending on the access points chosen for backscatter communication. First, we place two receiving APs between the transmitter and the tag and evaluate the beam-steering capability of the tag and also evaluate the effectiveness of the proposed algorithm to determine the best tag configuration. Figure\ref{fig:rssi_location} shows the RSSI of the received signal at two APs as a function of phase delay between clocks driving modulating consecutive antennas($\Delta$). As the value of $\Delta$ changes, the backscattered signal steers from AP4($\Delta =-100$) to AP5($\Delta = 145$).

\begin{figure}[t]
\centering
\begin{minipage}{0.23\textwidth}
\includegraphics[width=\linewidth]{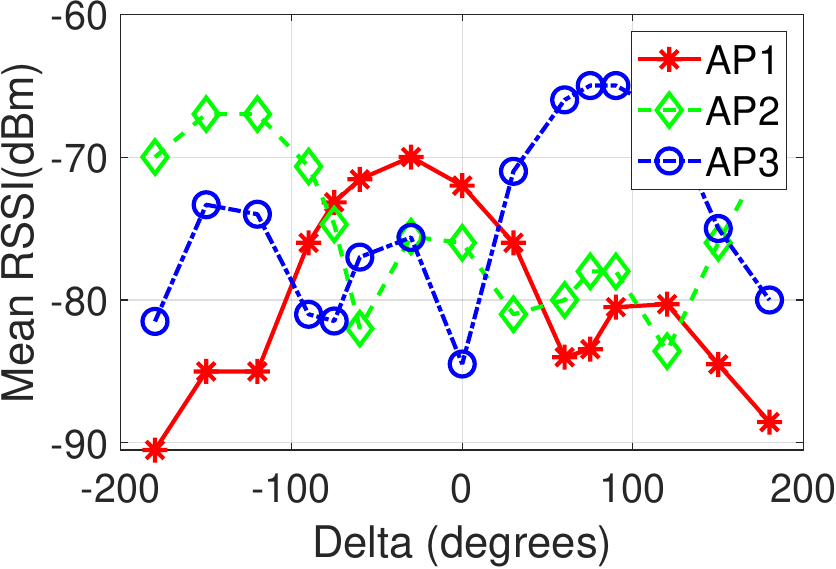}
\subcaption{Mean RSSI for Rx APs placed beyond the tag location}
\label{fig:rssi_0deg}
\end{minipage}
\begin{minipage}{0.23\textwidth}
\includegraphics[width=\linewidth]{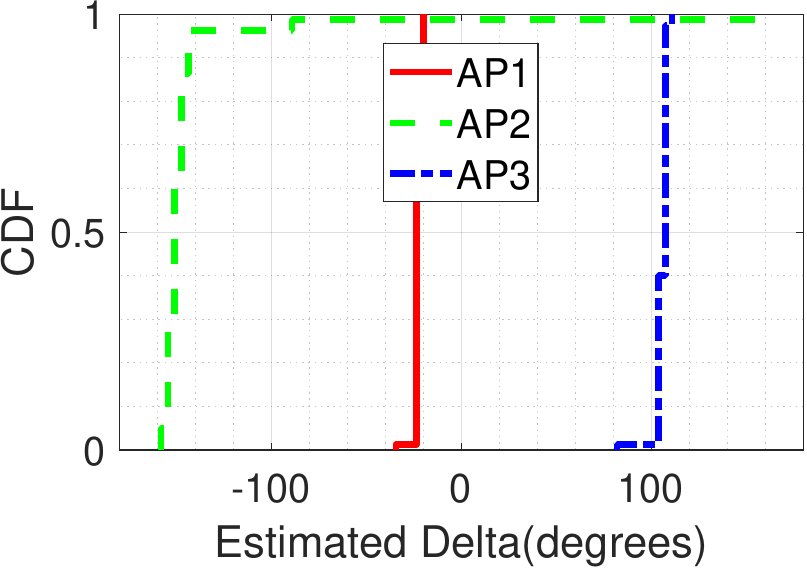}
\subcaption{CDF plot of estimated best tag phase setting}
\label{fig:Delta_0deg_anechoic}
\end{minipage}
\caption{Figure showing the algorithm's robustness to find the best phase setting tag for different Rx AP placement}
\end{figure}

Now, we place the receiving APs so the tag is between the transmitter and 3 receiver APs and repeat the experiments. In this case, AP1 is placed at 0-degree w.r.t. to the tag's normal, and AP2, AP3 are on either side of the normal. As shown in figure\ref{fig:rssi_0deg}, with changing $\Delta$, the backscattered signal power increases from AP1's direction to AP3's direction.Figure~\ref{fig:Delta_0deg_anechoic} shows the CDF of the estimated delta to steer the signal toward different APs. The estimated tag configuration closely matches the true values of $\Delta$ that would steer the signals towards each AP individually. The sharp CDFs indicate the algorithm can estimate the best Delta reliably to steer the backscattered signal to any desired receiving AP location.  

\begin{figure}[t]
\centering
\begin{minipage}{0.23\textwidth}
\includegraphics[width=\linewidth]{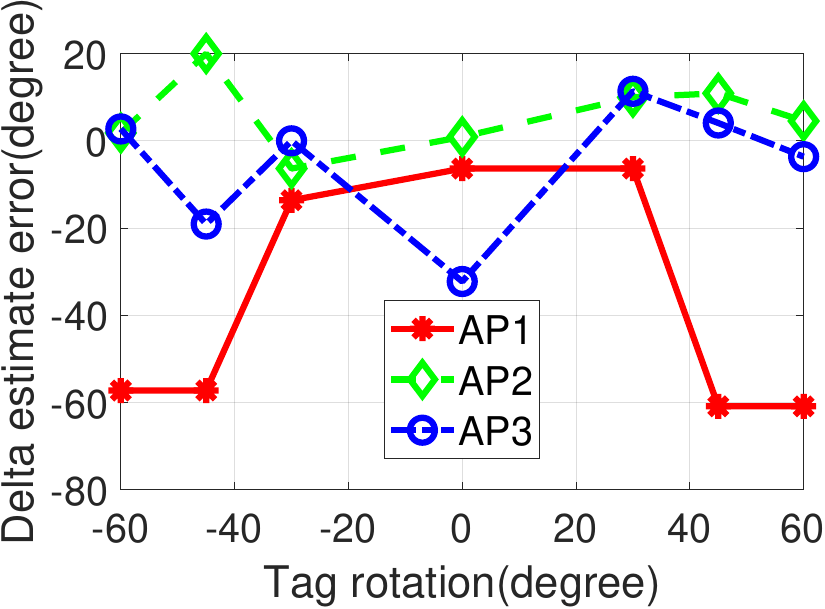}
\subcaption{Error in estimated delta for different APs as a function of tag rotation angle}
\label{fig:Delta_err}
\end{minipage}
\begin{minipage}{0.23\textwidth}
\includegraphics[width=\linewidth]{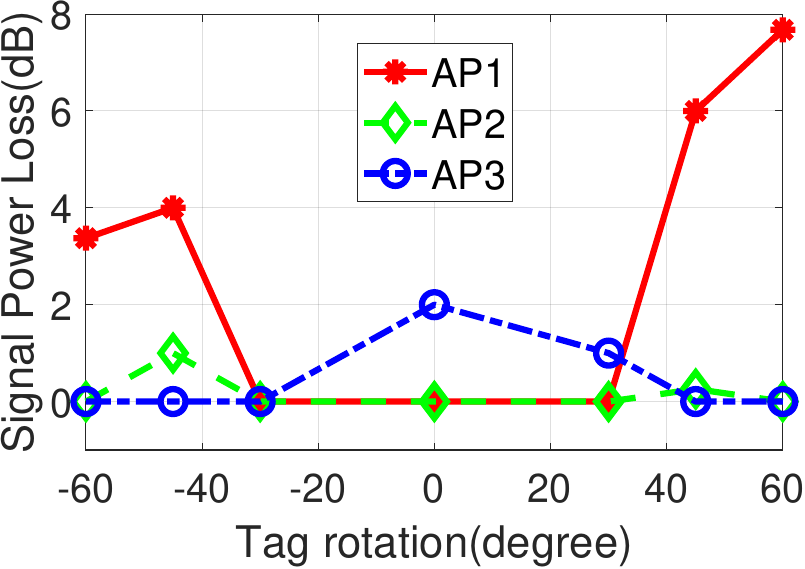}
\subcaption{Loss of signal power due to error in estimating tag's best phase setting}
\label{fig:rssi_err}
\end{minipage}
\caption{Algorithm's performance to estimate best phase setting for different tag rotation angles}
\end{figure}

\noindent
\textbf{Effect of Tag's rotation:}
So far in evaluations, we assumed the excitation signal is incident normal to the tag. But in reality, the tag can be oriented in any direction and it is possible to have oblique incidence of excitation signal.So, we maintain the receiver locations as in the previous case and retook the RSSI and CSI measurements by rotating the tag by different angles both clockwise and in anti-clockwise directions. Our goal here is to see if the backscattered signal power is maximum for a different value of $\Delta$ compared to the previous case and show that the algorithm is able to adapt in finding the best phase configuration for oblique signal incidence. Figure\ref{fig:Delta_err}  plots the error in the estimated $\Delta$ given by the algorithm and the true $\Delta$ that steers the signal towards a receiving AP as a function of tag's rotation.  For APs that are located on either side of tag's normal(AP2,AP3), the error in the estimated $\Delta$ is within $20^{\circ}$. For AP1, that lies on the straight line joining the Tx and the tag, we observe that as the tag rotation increased beyond $45^{\circ}$, the error in estimated $\Delta$ goes more than $20^{\circ}$. To understand the affect of the estimation error in the beamsteering gain, we plot the loss of signal power in figure\ref{fig:rssi_err} as a function of tag rotation angle. For AP1 that is in straight line with the tag, when the tag rotation is beyond 45 degree, there is upto 6 dB power loss. For AP2 and AP3, the estimation error leads to less than 2 dB loss of signal power.  

\begin{figure}[t]
\centering
\begin{minipage}{0.23\textwidth}
\includegraphics[width=\linewidth]{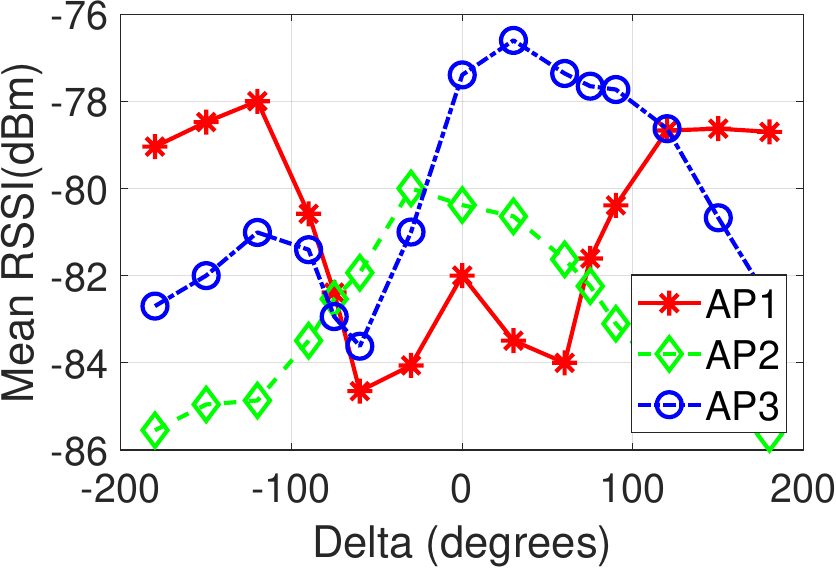}
\subcaption{Mean RSSI for different APs as a function of the tag's phase}
\label{fig:rssi_45deg_anticlock_outdoors}
\end{minipage}
\begin{minipage}{0.23\textwidth}
\includegraphics[width=\linewidth]{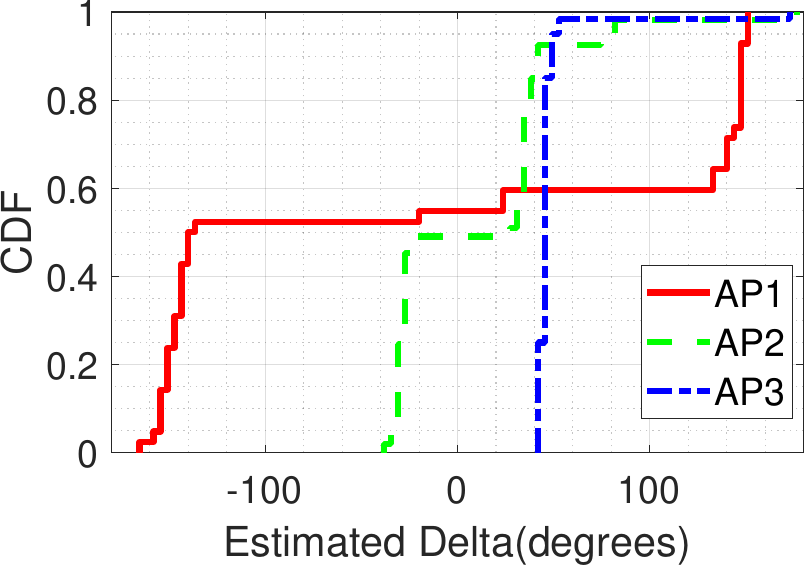}
\subcaption{CDF plot of estimated best tag phase setting for the APs.}
\label{fig:Delta_45deg_anticlock_outdoors}
\end{minipage}
\caption{Algorithm's performance in outdoor experiments with APs in multiple directions from the tag}
\end{figure}

\noindent
\textbf{Long range experiments:} We repeat the experiments outdoors in a 60 x 40m open area and place three receiving APs 10m from the tag in different directions and the transmitter is kept 10m from the tag and evaluate the tag's capability to steer the signal towards the access points. Figure\ref{fig:rssi_45deg_anticlock_outdoors} shows the RSSI at three APs as $\Delta$ on the tag is varied. Figure\ref{fig:Delta_45deg_anticlock_outdoors} shows the CDF of estimated $\Delta$ to steer the backscatter signal to each of the APs. In case of AP1, $\Delta$ ranging from $-180^{\circ}$ to $-150^{\circ}$ and $150^{\circ}$ to $180^{\circ}$, result in signal strength close to that of optimal phase setting. The CDF curves are also sharp in the same region indicating the proposed algorithm is able to find phase $\Delta$ to steer the signal towards the AP1 direction. Similarly, for AP2 the estimated phase $\Delta$ leads to 1 to 2 dB SNR loss from the optimal case. 

 \begin{figure}[!t]
 \centering
 \begin{minipage}{0.23\textwidth}
 \includegraphics[width=\linewidth]{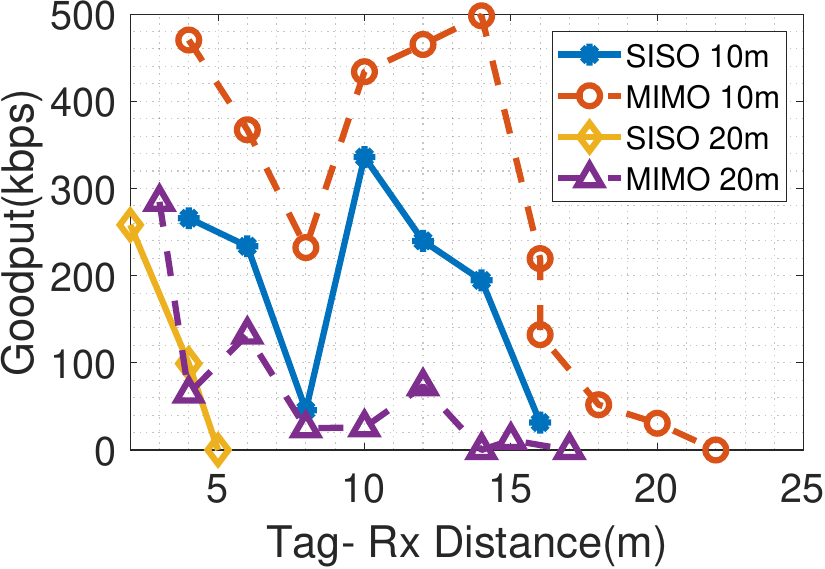}
 \subcaption{Goodput}
 \label{fig:Goodput}
\end{minipage}
 \begin{minipage}{0.23\textwidth}
 \includegraphics[width=\linewidth]{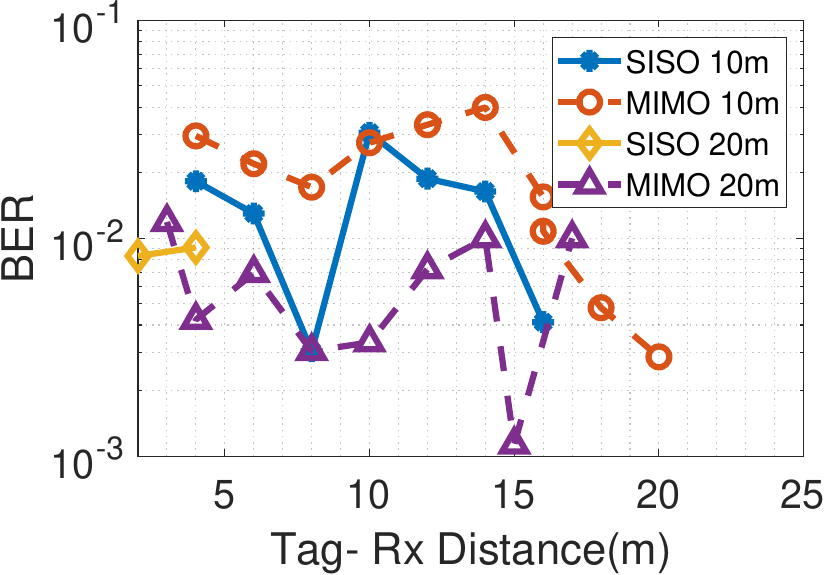}
 \subcaption{Bit error rate }
 \label{fig:BER}
 \end{minipage}
 \caption{Comparison of End to End goodput and bit error rate performance for single antenna and \name tag }
\end{figure}

%  \begin{figure}[!t]
%  \centering
%  \includegraphics[width=.9\linewidth]{figures/rssi_cdf_0deg_oracle.pdf}
%  \caption{Comparison of the CDFs of the RSSI for the estimated best phase setting and the true best phase setting (oracle) when the Tx, tag and the Rx are in a straight line.}
%  \label{fig:rssi_cdf_0deg}
% \end{figure}

%  \begin{figure}[!t]
%  \centering
%  \includegraphics[width=.9\linewidth]{figures/rssi_cdf_45deg_oracle.pdf}
%  \caption{Comparison of the CDFs of the RSSI for the estimated best phase setting and the true best phase setting (oracle) when the Rx is kept 45 degree away from the tag's normal.}
%  \label{fig:rssi_cdf_45deg}
% \end{figure}

%  \begin{figure}[!t]
%  \centering
%  \includegraphics[width=.9\linewidth]{figures/Goodput_crop.pdf}
%  \caption{Goodput for single antenna vs multi-antenna tag as a function of Tag to Rx separation for different Tx to Tag separations.}
%  \label{fig:Goodput}
% \end{figure}
%  \begin{figure}[!t]
%  \centering
%  \includegraphics[width=.9\linewidth]{figures/BER_crop.pdf}
%  \caption{BER for single antenna vs multi-antenna tag as a function of Tag to Rx separation for different Tx to Tag separations.}
%  \label{fig:BER}
% \end{figure}

%  \subsection{Show the improvement in SNR with the estimated best phase settings on the tag}
 
%  CDF plots of RSSI for the 4 measured configurations and also CDF plot to show the improvement in the RSSI with the estimated best configuration.

\subsection{End-to-End Experiments}
\noindent
In this section, we will present the end-to-end throughput and BER results for different Tx, Rx and tag locations. To measure the throughput and BER of the tag data for a given arrangement of Tx AP, tag and the Rx AP, we set the tag in the best-estimated phase configuration. Then the Tx AP sends 802.11b WiFi packets on channel 1(2412 MHz) at 24 dBm average power and the tag embeds its data on top of the incident WiFi packet and shifts it to channel 6 (2437MHz). We run Wireshark on the Rx AP to analyze the packets received by it and process them to compute the Goodput \\(accounting for the overhead due to wakeup pattern in the Transmitted waveform) and the bit error rate for different Tag to Rx distances. Finally, we compare the performance of single antenna backscatter tag and \name tag to show the improvements. Note that, we chose single antenna tag as a baseline instead of a high gain antenna array because the high gain antenna array directs the backscattered signal only in a fixed direction and hence it extends the range, only in that fixed direction.

\textbf{Goodput}: Figure\ref{fig:Goodput} plots the Goodput in kbps for two cases: a) Tag is placed 10m away from the transmitter and b) Tag is placed 20m away from the transmitter. We observe that the Goodput in general decreases with the increase of Tag to Rx distance due to an increase in path loss. A key point to note here is that the Goodput for \name tag is always higher than a single antenna tag and \name tag achieves an average of 400 kbps rate up to 15m Tag to Rx separation. For the first case, \name works up to 22m whereas single antenna tag works only up to 14m. An anomaly is observed in both single and multi-antenna cases when Tag to Rx is 8m because of destructive interference of multipath at that position. 
For the second case of 20m Tx to Tag separation, \name tag works for a range of up to 14m whereas single antenna tags work only up to 4m. 

\textbf{Bit Error Rate}: Figure\ref{fig:BER} plots the BER for different Tag to Rx separations for 10m Tx to tag separation and 20m Tx to tag separation. We note that BER in case of \name tag stays at around $10^{-2}$ for both 10 m and 20m Tx to tag separations.

\subsection{Range of \name tag:}
\begin{figure}[!t]
 \centering
 \includegraphics[width=.9\linewidth]{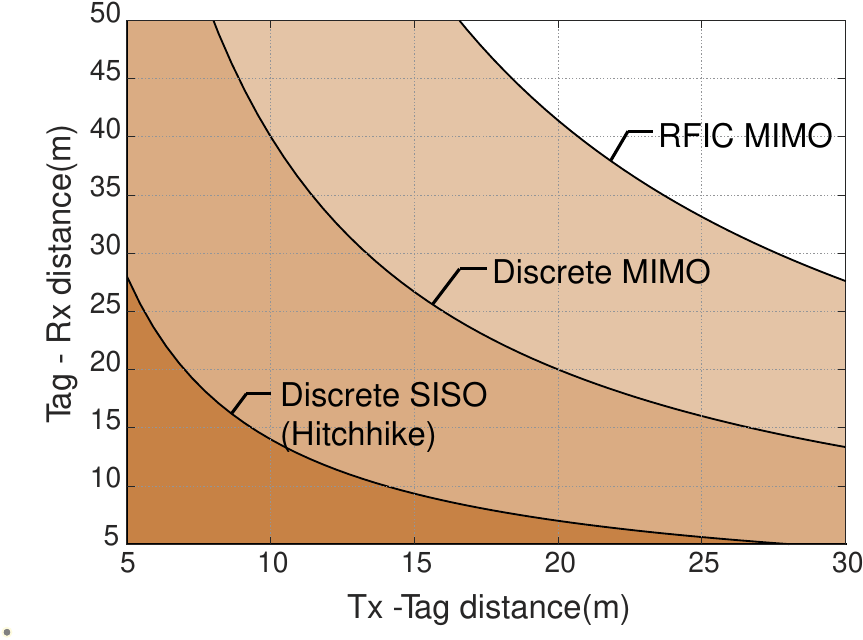}
 \caption{Tx to Tag separation vs Tag to Rx separation for single antenna and \name tags. }
 \label{fig:D1D2}
\end{figure}

\noindent
Here we describe the range of \name tag as a function of Tx to tag distance as plotted in Figure\ref{fig:D1D2}. In case of single antenna tag, d1$\times$d2 $\le$ 140$m^{2}$ and for \name tag, d1$\times$d2 $\le$ 400$m^{2}$. Any tag location satisfying these constraints is a feasible location and we see that the \name tag has 20m Tx to Tag and 20m Tag to Rx separation as a feasible point. So, \name tag can be easily supported in a network of access points that are separated by 40m.

% \subsection{Mobility experiments:}
 
% \subsection{Two tag experiments:}
% Setup two tags with different programmed wake-up patterns and do the estimation of best phase settings for each tag simultaneously. Finally show that the signal strength increased for the estimated tag configurations. 

\subsection{RFIC Tag vs Discrete PCB Version}
\noindent
Table \ref{tab: tab1} shows the comparison of \name's tag RFIC and discrete component version, under different figure of merits including the wake-up receiver sensitivity (at $1e-3$ missed detection rate), the insertion loss of the switch, and key differences in the backscattering performance. Furthermore, as expected the \name's RFIC tag has around 500 times of power reduction compared to discrete component one, and more compact size in a single 1.7mm $\times$ 1.2mm die instead of having 10 discrete ICs. The RFIC tag also shows a longer range of d1$\times$d2 $\le$ 828$m^{2}$ in MIMO mode. However, the discrete component version has more flexibility to generate different clock frequencies with multi-phased clocks on FPGA, whereas the RFIC tag has fixed frequency clock generation. 

\begin{table}
\begin{tabular}{ |c|c|c|p{2cm}|p{2cm}| } 
 \hline
 & Power & d1$\times$d2 range & WuRx sensitivity &  Insertion Loss\\ \hline
RFIC & 88$\mu$W & 828$m^{2}$ & -43.4 dBm & 7.8 dB\\
Discrete & 40mW & 400$m^{2}$ & -38.5dBm  &  9 dB\\ 
 \hline
\end{tabular}
 \caption{Table comparing the RFIC and the Discrete component \name tag}
 \label{tab: tab1}
\end{table}

%!TEX root = main.tex

\section{Related Work}
%\todo{Time Modulated array: It only synthesizes a beampattern by dutycycling the antennas. We also have frequency shifting capability.}

\noindent
\name is related to prior backscatter-based works\cite{hitchhike,ISSCC-backscatter,dunna2021syncscatter,interscatter,shyambackscatter,freerider,wifibackscatter,passive-wifi,TMA} that make use of ambient wireless signals to enable low-power communications. However, most of the prior backscatter works are based on a single antenna on the backscatter tags. 
%We differ from other backscatter works in the following aspects:

\noindent\textbf{Time Modulated Arrays:} Our work is closely related to time modulated arrays\cite{TMA,maneiro2017time,wang2012time} where an additional time dimension is introduced in an antenna array by controlling the time duration for which signal is captured at each antenna using RF-switches. They are primarily used at receivers to create beampatterns with ultra-low sidelobes. A byproduct of duty-cycling the antennas on the array result in harmonics that are unwanted and are kept very low for Electromagnetic compatibility. Our work on the other hand, uses the antenna arrays at an intermediate backscatter node to reflect the signals by maximizing the energy of the harmonics in the reflections.

\noindent\textbf{Retro-reflective backscatter:} Recent works \cite{meng202112,trotter2012multi} propose the use of multiple antennas on the Wi-Fi and RFID backscatter tags based on Van-Atta arrays\cite{sharp1960van} to reflect the incident signals back to the same location as the transmitter and provides beamforming gain in the process. Using this in the context of WiFi requires co-located Tx and Rx access points(not available commercially). They also need strong RF channel filters at the transmitter and receiver to keep the out-of-band emissions happening at the transmitter below the retro-reflected signal power, otherwise, the range of the tag will be severely limited. Our work, on the other hand, steers the incident signal on the tag towards another AP and do not suffer from the transmitter's out-of-band leakage. 

\noindent\textbf{MIMO backscatter:} MOXcatter\cite{MOXcatter} introduces to use a dedicated spatial stream as an excitation signal to a single antenna tag by using a multi-antenna 802.11n transmitter. While using both spatial streams, \cite{MOXcatter} modulates only one bit per whole packet and leads to bit rate of 1kbps. VMscatter\cite{liu2020vmscatter} builds on top of this and uses multiple antennas at the backscatter tag to encode tag data using space time codes. Space time codes offer diversity gain and so effectively backscatter data rate remains to be the same with a reduced bit error rate. Their system needs IQ sample data from the receivers to decode tag data and hence it is not compatible with COTS WiFi devices. \name tag on the other hand uses COTS devices and can leverage doing higher order modulation\cite{yuan2022high} like QPSK and CCK modulation to achieve higher data rate in 11b standard because of the available increase in signal power due to beamforming gain at the tag.

%For example, consider Alamouti code(STBC code with the highest code rate 1) sends two different symbols at different antennas in the first time slot but they repeat the same set of symbols by conjugating and sending them in the next slot as well resulting in sending only 2 symbols in 2 time slots.

\noindent\textbf{Tunnel diode backscatter:} Tunnelscatter\cite{varshney2019tunnelscatter,amato2018tunneling} uses a tunnel diode in the negative resistance region to create a reflection coefficient more than 1 thus amplifying the reflected signal from the tag. However, this approach have limited amplification bandwidth and is not suitable for WiFi signals. \cite{bidirectional_amplifier} uses a set of bidirectional amplifiers in a Van Atta array which have a special property to amplify the signals entering both the amplifier input and output ports and retro-reflect the incident signals. Even with increase in tag's reflected power, Van atta array tags are limited by the self-interference caused by transmitter. \name tag on the other hand deals with the self-interference issue by beamsteering the tag's signal to entirely different AP location.
%\noindent\textbf{Backscatter Networking:} 
%\vspace{-0.3 cm}
\section{Discussion and Future work:}
\name tag provides a long backscatter range using multiple antennas on the tag. Using multiple antennas on the tag increases the size of tag PCB. For some applications like air quality sensing,humidity sensing and security camera, the tag size is not a concern and can be mounted on walls next to the sensor. To reduce form factor of the tag, one can use low profile antennas like chip antennas or patch antennas\cite{chip_antenna} on the PCB and we leave this design as a future work. 
%!TEX root = main.tex 
%\vspace{-5 mm}
%\section{Conclusion and Discussion}\label{sec:conclusion}

% \section{Conclusion}

% In this paper, we have discussed the design of a backscatter tag that can redirect the incident signals from one AP to another AP to motivate a wide-range deployment of backscatter tags in a mesh network of WiFi Access points. We also presented a protocol of how the tag would interact with the AP using a wake-up receiver and outlined an algorithm to find the optimal phase shifts on the tag to increase the working range of Tag from the access points. Finally, we presented beam-pattern measurement results from the backscatter tag and end to end results in terms of throughput and BER.

% \newpage
\clearpage
\bibliographystyle{abbrv}
\balance
\bibliography{references}

\begin{thebibliography}{10}

\bibitem{chip_antenna}
2.4ghz/5ghz chip antenna.
\newblock
  \url{https://www.digikey.com/en/products/detail/molex/1461750001/6071805}.

\bibitem{LT5534}
50mhz to 3ghz envelope detector lt5534.
\newblock
  \url{https://www.analog.com/media/en/technical-documentation/data-sheets/5534fc.pdf}.

\bibitem{rfswitch}
Adg902 rf switches analog devices.
\newblock \url{https://www.analog.com/en/products/adg902.html}.

\bibitem{HMC247}
Analog phase shifter.
\newblock
  \url{https://www.analog.com/media/en/technical-documentation/data-sheets/hmc247.pdf}.

\bibitem{cmod}
Cmod a7: Breadboardable artix-7 35t fpga module.
\newblock
  \url{https://digilent.com/shop/cmod-a7-breadboardable-artix-7-fpga-module/}.

\bibitem{WiFi_surveys}
Iot device specifications.
\newblock
  \url{https://www.netspotapp.com/wifi-site-survey/wifi-planning-and-site-survey.html}.

\bibitem{MAX49140}
Ultra low power high speed comparator.
\newblock
  \url{https://www.maximintegrated.com/en/products/analog/amplifiers/MAX49140.html#online-ds}.

\bibitem{amato2018tunneling}
F.~Amato, C.~W. Peterson, B.~P. Degnan, and G.~D. Durgin.
\newblock Tunneling rfid tags for long-range and low-power microwave
  applications.
\newblock {\em IEEE Journal of Radio Frequency Identification}, 2(2):93--103,
  2018.

\bibitem{RF_switch}
A.~devices.
\newblock Adg 902 rf switch from analog devices for 2.4ghz band.
\newblock
  \url{https://www.analog.com/media/en/technical-documentation/data-sheets/ADG901_902.pdf}.

\bibitem{dunna2021syncscatter}
M.~Dunna, M.~Meng, P.-H. Wang, C.~Zhang, P.~P. Mercier, and D.~Bharadia.
\newblock Syncscatter: Enabling wifi like synchronization and range for wifi
  backscatter communication.
\newblock In {\em NSDI}, pages 923--937, 2021.

\bibitem{ISSCC22-Shihkai}
A.~et.
\newblock {Anonymous to preserve double blind review}, 2022.

\bibitem{bidirectional_amplifier}
F.~Farzami, S.~Khaledian, B.~Smida, and D.~Erricolo.
\newblock Reconfigurable dual-band bidirectional reflection amplifier with
  applications in van atta array.
\newblock {\em IEEE Transactions on Microwave Theory and Techniques},
  65(11):4198--4207, 2017.

\bibitem{nexmoncsi}
F.~Gringoli, M.~Schulz, J.~Link, and M.~Hollick.
\newblock Free your csi: A channel state information extraction platform for
  modern wi-fi chipsets.
\newblock pages 21--28, 10 2019.

\bibitem{TMA}
C.~He, L.~Wang, J.~Chen, and R.~Jin.
\newblock Time-modulated arrays: A four-dimensional antenna array controlled by
  switches.
\newblock {\em Journal of Communications and Information Networks}, 3(1):1--14,
  2018.

\bibitem{interscatter}
V.~Iyer, V.~Talla, B.~Kellogg, S.~Gollakota, and J.~Smith.
\newblock Inter-technology backscatter: Towards internet connectivity for
  implanted devices.
\newblock In {\em {SIGCOMM}}, 2016.

\bibitem{far-field}
R.~C. Johnson, H.~A. Ecker, and J.~S. Hollis.
\newblock Determination of far-field antenna patterns from near-field
  measurements.
\newblock {\em Proceedings of the IEEE}, 61(12):1668--1694, 1973.

\bibitem{wifibackscatter}
B.~Kellogg, A.~Parks, S.~Gollakota, J.~R. Smith, and D.~Wetherall.
\newblock Wi-fi backscatter: Internet connectivity for rf-powered devices.
\newblock In {\em ACM SIGCOMM Computer Communication Review}, 2014.

\bibitem{passive-wifi}
B.~Kellogg, V.~Talla, S.~Gollakota, and J.~R. Smith.
\newblock Passive wi-fi: Bringing low power to wi-fi transmissions.
\newblock In {\em Proceedings of the 13th Usenix Conference on Networked
  Systems Design and Implementation}, NSDI'16, page 151–164, USA, 2016.
  USENIX Association.

\bibitem{liu2020vmscatter}
X.~Liu, Z.~Chi, W.~Wang, Y.~Yao, and T.~Zhu.
\newblock Vmscatter: A versatile $\{$MIMO$\}$ backscatter.
\newblock In {\em 17th $\{$USENIX$\}$ Symposium on Networked Systems Design and
  Implementation ($\{$NSDI$\}$ 20)}, pages 895--909, 2020.

\bibitem{phased_array_book}
R.~J. Mailloux.
\newblock {\em Phased Array Antenna Handbook}.
\newblock Artech House, Inc., USA, 3rd edition, 2017.

\bibitem{maneiro2017time}
R.~Maneiro-Catoira, J.~Br{\'e}gains, J.~A. Garc{\'\i}a-Naya, and L.~Castedo.
\newblock Time modulated arrays: From their origin to their utilization in
  wireless communication systems.
\newblock {\em Sensors}, 17(3):590, 2017.

\bibitem{meng202112}
M.~Meng, M.~Dunna, H.~Yu, S.~Kuo, H.~P-Wang, D.~Bharadia, and P.~P. Mercier.
\newblock 12.2 improving the range of wifi backscatter via a passive
  retro-reflective single-side-band-modulating mimo array and non-absorbing
  termination.
\newblock In {\em 2021 IEEE International Solid-State Circuits Conference
  (ISSCC)}, volume~64, pages 202--204. IEEE, 2021.

\bibitem{sharp1960van}
E.~Sharp and M.~Diab.
\newblock Van atta reflector array.
\newblock {\em IRE Transactions on Antennas and Propagation}, 8(4):436--438,
  1960.

\bibitem{trotter2012multi}
M.~S. Trotter, C.~R. Valenta, G.~A. Koo, B.~R. Marshall, and G.~D. Durgin.
\newblock Multi-antenna techniques for enabling passive rfid tags and sensors
  at microwave frequencies.
\newblock In {\em 2012 IEEE International Conference on RFID (RFID)}, pages
  1--7. IEEE, 2012.

\bibitem{varshney2019tunnelscatter}
A.~Varshney, A.~Soleiman, and T.~Voigt.
\newblock Tunnelscatter: Low power communication for sensor tags using tunnel
  diodes.
\newblock In {\em The 25th Annual International Conference on Mobile Computing
  and Networking}, pages 1--17, 2019.

\bibitem{shyambackscatter}
A.~Wang, V.~Iyer, V.~Talla, J.~R. Smith, and S.~Gollakota.
\newblock Fm backscatter: Enabling connected cities and smart fabrics.
\newblock In {\em NSDI}, pages 243--258, 2017.

\bibitem{ISSCC-backscatter}
P.-H.~P. Wang, C.~Zhang, H.~Yang, D.~Bharadia, and P.~P. Mercier.
\newblock 20.1 a 28$\mu$w iot tag that can communicate with commodity wifi
  transceivers via a single-side-band qpsk backscatter communication technique.
\newblock In {\em 2020 IEEE International Solid-State Circuits
  Conference-(ISSCC)}, pages 312--314. IEEE, 2020.

\bibitem{wang2012time}
Y.~Wang and A.~Tennant.
\newblock Time-modulated reflector array.
\newblock {\em Electronics Letters}, 48(16):972--974, 2012.

\bibitem{yuan2022high}
L.~Yuan and W.~Gong.
\newblock High-throughput backscatter using commodity wifi.
\newblock In {\em Proceedings of the 20th Annual International Conference on
  Mobile Systems, Applications and Services}, pages 535--536, 2022.

\bibitem{hitchhike}
P.~Zhang, D.~Bharadia, K.~Joshi, and S.~Katti.
\newblock Hitchhike: Practical backscatter using commodity wifi.
\newblock In {\em {SenSys}}, 2016.

\bibitem{freerider}
P.~Zhang, C.~Josephson, D.~Bharadia, and S.~Katti.
\newblock Freerider: Backscatter communication using commodity radios.
\newblock In {\em Proceedings of the 13th International Conference on Emerging
  Networking EXperiments and Technologies}, CoNEXT '17, page 389–401, New
  York, NY, USA, 2017. Association for Computing Machinery.

\bibitem{MOXcatter}
J.~Zhao, W.~Gong, and J.~Liu.
\newblock Spatial stream backscatter using commodity wifi.
\newblock In {\em Proceedings of the 16th Annual International Conference on
  Mobile Systems, Applications, and Services}, pages 191--203, 2018.

\end{thebibliography}
\end{document}